\documentclass[aps,pra,showpacs,superscriptaddress,groupedaddress]{revtex4}  

\usepackage{graphicx}   
\usepackage{bm}             
\usepackage{amssymb}   
\usepackage{epsfig}
\usepackage{color}

\begin{document}


\title{Vibrational excitation cross sections for non-equilibrium nitric oxide-containing plasma}

\author{V.~Laporta}
\email{vincenzo.laporta@istp.cnr.it}
\affiliation{Istituto per la Scienza e Tecnologia dei Plasmi, CNR, 70126 Bari, Italy}

\author{L.~Vialetto}
\affiliation{Theoretical Electrical Engineering, Faculty of Engineering, Kiel University, Kaiserstra\ss e 2, 24143 Kiel, Germany}
\affiliation{Department of Applied Physics, Eindhoven University of Technology, P.O. Box 513, 5600 MB Eindhoven, the Netherlands}

\author{V.~Guerra}
\affiliation{Instituto de Plasmas e Fus\~{a}o Nuclear, Instituto Superior Tecnico, Universidade de Lisboa, 1049-001 Lisboa, Portugal}

\begin{abstract}
A full set of vibrationally-resolved cross sections for electron impact excitation of NO(X\,$^2\Pi, v$) molecules is calculated from \textit{ab initio} molecular dynamics, in the framework of the Local-Complex-Potential approach. Electron-vibration energy exchanges in non-equilibrium thermodynamic conditions are studied from a state-to-state model accounting for all electron impact excitation and de-excitation processes of the nitric oxide vibration manifold, and it is shown that the calculated vibration relaxation times are in good agreement with the experimental data. The new vibrational excitation cross sections are used in a complete electron impact cross section set in order to obtain non-equilibrium electron energy distributions functions and to calculate electron transport parameters in NO. It is verified that the new cross sections bring a significant improvement between simulations and experimental swarm data, providing an additional validation of the calculations, when used within the complete set of cross sections investigated in this work. 
\end{abstract}

\pacs{xxxxx}

\maketitle

\section{\label{sec:intro} Introduction}

Nitrogen molecule (N$_2$) and its oxide compounds (NO$_x$), in particular nitric oxide (NO), comprise a significant fraction of the Earth’s atmosphere and they are of primary importance for life on Earth and for many human activities. The vibrationally excited levels of NO, for example, play a fundamental role in understanding the chemical processes occurring in the upper atmosphere. In fact, it has been proven that their distribution is very sensitive to the fast auroral electrons in ionosphere, which are directly connected to the solar activity~\cite{BOUZIANE2022905}. As a second example, the vibrational excitation of NO molecule is also very relevant in hypersonic applications and in the atmospheric reentry problem of shuttle spacecrafts, where it is related to the contribution to radiation heating~\cite{SAHAI2020106752, :/content/aip/journal/jap/118/13/10.1063/1.4931769, doi:10.1063/1.4804388}.

From the biological side, NO molecule is a metabolic nitrogen compound in gas form, and it plays an important role in the physiological regulation of the cardiovasculature in the human body~\cite{NASEEM200533, Feng:2012wq}. Moreover, in novel generation of biosensors for real-time bioimaging, NO is a key signaling molecule for vasodilatation and neurotransmission at low concentrations and a defensive cytotoxin at higher concentrations~\cite{https://doi.org/10.1038/sj.bjp.0706458, Jeon:2014tu}. As another example, from a technological point of view, nitrogen fixation processes, which refer to the formation of compounds such as ammonia and nitrate from nitrogen molecules, is one of the most important industrial activities in agriculture~\cite{Chen:2021tk}. The key point is that current ammonia synthesis methods make use of large amount of fossil fuels, which contribute to the global current human CO$_2$ emissions. In this respect, a worldwide interest about the possibility to explore sustainable nitrogen fixation processes by using cold plasmas has been growing in the last years and new green technologies have been proposed very recently~\cite{doi:10.1021/acssuschemeng.0c01815, D0EE03763J, D0GC03521A}.

In order to study the transport properties and chemistry of NO-containing systems, several kinetic models appeared recently in literature~\cite{Bahnamiri_2021, doi:10.1098/rsta.2014.0331, BOUZIANE2022905} and, at same time, new sets of cross sections and rate coefficients for chemical reactions for the NO molecule have been proposed~\cite{doi:10.1029/2003GL019151, doi:10.1063/1.5114722, doi:10.1063/1.4961372}, including for heavy particle collisions~\cite{ARMENISE2021111325, doi:10.1021/acs.jpca.0c09999}. To this purpose, in this work, we present a novel study of NO molecules interacting with electrons, covering \textit{ab initio} molecular dynamics and non-equilibrium kinetics modelling.

The paper is organized as follows. In sections \ref{sec:th} and \ref{sec:results}, we use the theoretical framework employed in recent papers~\cite{10.1088/1361-6595/ab86d8, Laporta_2020} to calculate dissociative cross sections by electron-impact for NO molecule, to extend the calculations to the process of vibrational excitation (VE) \textit{i.e.}:
\begin{equation}
e(\epsilon)  + \mathrm{NO}(\mathrm{X}\,^2\Pi; v) \to \mathrm{NO}^- \to e + \mathrm{NO}(\mathrm{X}\,^2\Pi;v')\,, \qquad v,v'=0,\ldots 53\,; \label{eq:VEprocess}
\end{equation}
a comparison with data available in literature and an update of the data presented in the Ref.~\cite{0963-0252-21-5-055018} is also included. In section~\ref{sec:relaxvib}, we discuss a 0D model for the vibrational relaxation of NO molecules in chemical non-equilibrium conditions when electrons are described by a Maxwellian distribution. The non-equilibrium effects for electrons are taken into account in section \ref{sec:EEDF}. Finally, section~\ref{sec:conc} summarizes the main conclusions and closes the paper.

\section{\label{sec:th} Theoretical model}

We consider collision energies where numerous NO$^-$ resonances exist and the VE reaction in (\ref{eq:VEprocess}) is dominated by resonant processes~\cite{0034-4885-31-2-302, Domcke199197}. Our aim is to cover a large range of incident electron energies, so we take into account five resonance states of NO$^-$:  the three low-lying states of  $^3\Sigma^-$, $^1\Sigma^+$ and $^1\Delta$ symmetries and two higher ones, with $^3\Pi$ and $^1\Pi$ symmetry, which lie close to the NO dissociation threshold. In the following, we number these resonances by $r=1,\ldots,5$, respectively.

In the Local-Complex-Potential (LCP) model, the VE cross section for the process in (\ref{eq:VEprocess}) -- for a NO molecule initially in vibrational level $v$ colliding with an electron of energy $\epsilon$ and ending to the final vibrational level $v'$ -- is given by~\cite{PhysRevA.20.194}:
\begin{equation}
\sigma_{v\to v'}(\epsilon) = \sum_{r=1}^5 \frac{2S_r+1}{(2S+1)\,2} \frac{g_r}{g\,2} 
\frac{64\,\pi^5\,m^2}{\hslash^4} 
\frac{k'}{k}\left|\langle 
\chi_{v'}|\mathcal{V}_r|\xi^r_v \rangle\right|^2\,, \label{eq:DExsec} 
\end{equation}
where $2S_r+1$ and $2S+1$ are the spin-multiplicities of the resonant anion state and of the neutral target state respectively, $g_r$ and $g$ represent the corresponding degeneracy factors, $k$($k'$) is the incoming (outgoing) electron momenta, $m$ is the electron mass, $\chi_{v'}$ stands for the final vibrational wave function of NO, $\mathcal{V}_r$ is the resonance coupling to the target state and $\xi^r_v$ is the resonance  wave function related to the initial vibrational wave function of NO.

For the sake of brevity,  in the present work we skip the full theoretical details on LCP model,  which can be found in Refs.~\cite{10.1088/1361-6595/ab86d8, Laporta_2020} and references therein. The only difference, in the theoretical approach, relies in suppressing here of the penetration factor $f_r$ given in Eq.~(6) of Ref.~\cite{10.1088/1361-6595/ab86d8} appearing in the coupling $\mathcal{V}_r$. The penetration function has been introduced in the paper~\cite{0963-0252-21-5-055018} as an \textit{ad hoc} factor but, the use of a penetration factor is an approximation that forces a wrong behavior at low energies ($\lesssim 0.1$ eV). Moreover, as discussed in the next Section, the cross sections obtained by neglecting this factor are in good agreement with experimental data. Note that the penetration factor does not affect the results for dissociative excitation and dissociative attachment presented in the papers in Refs.~\cite{10.1088/1361-6595/ab86d8, Laporta_2020}.

\begin{figure}
\centering
\includegraphics[scale=.33]{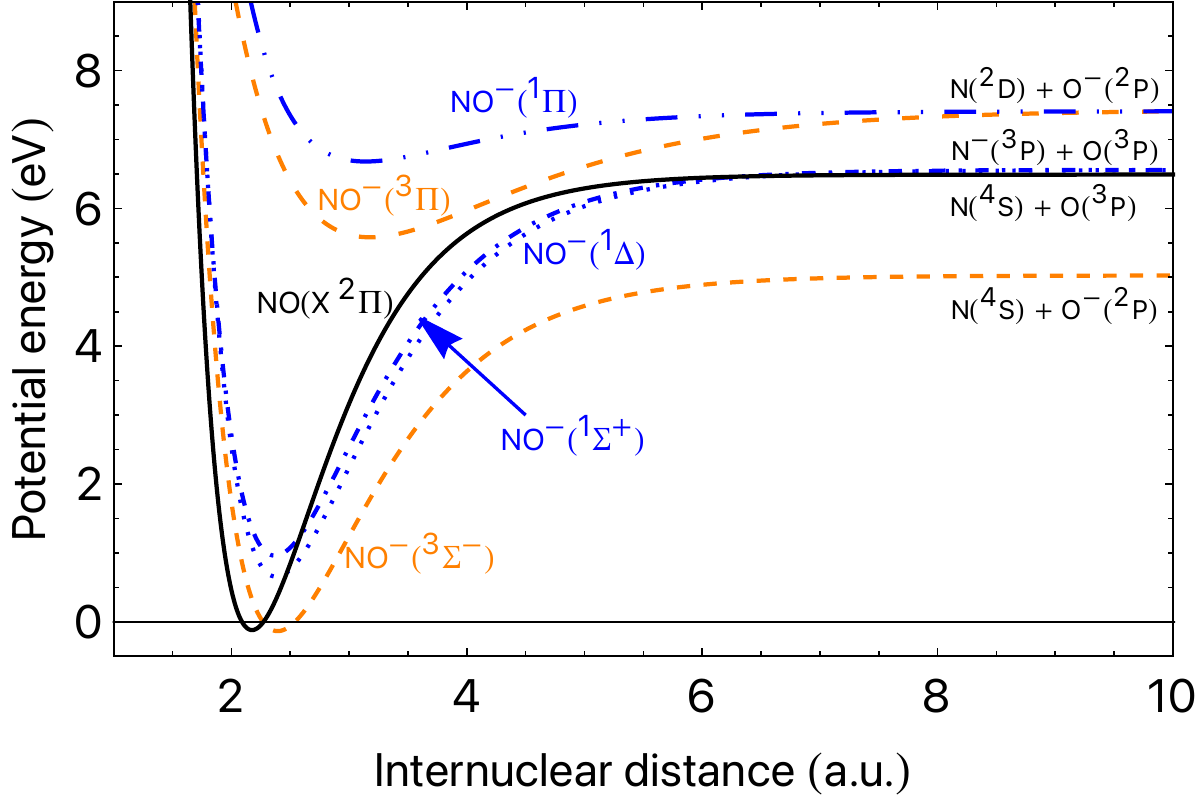} \hspace{.5cm} \includegraphics[scale=.34]{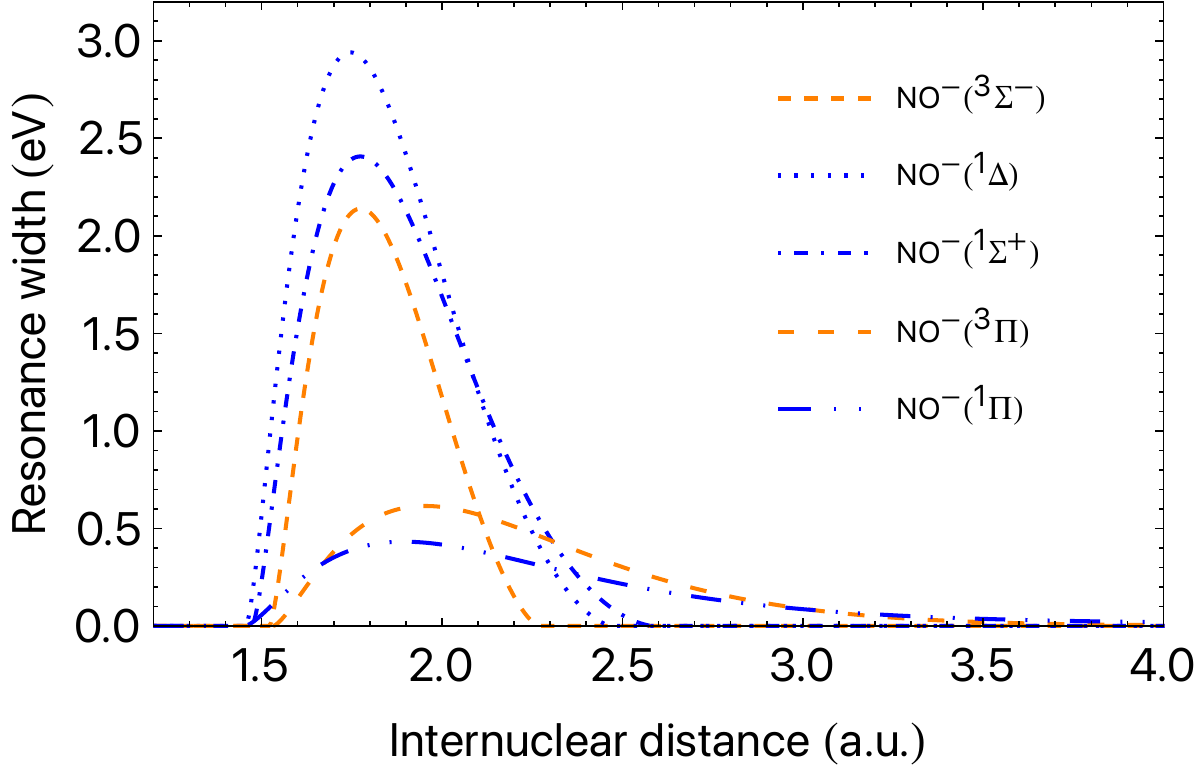} 
\caption{Molecular data involved in the calculations as determined in the article in Ref.~\cite{10.1088/1361-6595/ab86d8}. Left: Potential energy curves for the ground electronic state of NO molecule (solid line) and for the five NO$^-$ resonances (broken lines) for rotational quantum number $j=0$.  Right: The corresponding widths of the five resonances. \label{fig:NOpot}}
\end{figure}

In Figure~\ref{fig:NOpot} we show the complete set of molecular data used in the calculations, \textit{i.e.} the potential energy curves for NO, in its ground electronic state $\mathrm{X}\,^2\Pi$ and for the five NO$^-$ resonant states, and autoionization widths for the resonances.  All potentials refer to the rotational quantum number $j=0$. For the energies considered here, molecular rotation can be considered at thermodynamic equilibrium and, accordingly, in the following rotational effects will be neglected. In Table~\ref{tab:NOviblev}  we report the list of the vibrational levels $v$ supported by the NO molecule calculated with the potential energy curves of Figure~\ref{fig:NOpot}. As a matter of fact, the present paper and Ref.~\cite{0963-0252-21-5-055018} share the same potential energy for the ground state of NO and hence the resulting vibrational levels are the same.

\begin{table}
\centering
\begin{tabular}{cccccc}
\hline
~~~$v$~~~ & $~~\epsilon_{v}$~(eV)~~ & ~~~$v$~~~ & $~~\epsilon_{v}$~(eV)~~& ~~~$v$~~~ & $~~\epsilon_{v}$~(eV)~~\\
\hline
  0  &     0.000  &   18  &     3.581  &   36  &     5.745 \\ 
  1  &     0.236  &   19  &     3.739  &   37  &     5.823 \\ 
  2  &     0.468  &   20  &     3.892  &   38  &     5.897 \\ 
  3  &     0.695  &   21  &     4.040  &   39  &     5.967 \\ 
  4  &     0.918  &   22  &     4.185  &   40  &     6.033 \\ 
  5  &     1.137  &   23  &     4.324  &   41  &     6.094 \\ 
  6  &     1.351  &   24  &     4.460  &   42  &     6.150 \\ 
  7  &     1.561  &   25  &     4.591  &   43  &     6.203 \\ 
  8  &     1.767  &   26  &     4.718  &   44  &     6.251 \\ 
  9  &     1.968  &   27  &     4.840  &   45  &     6.294 \\ 
10  &     2.164  &   28  &     4.958  &   46  &     6.333 \\ 
11  &     2.357  &   29  &     5.072  &   47  &     6.368 \\ 
12  &     2.545  &   30  &     5.181  &   48  &     6.399 \\ 
13  &     2.729  &   31  &     5.286  &   49  &     6.425 \\ 
14  &     2.908  &   32  &     5.386  &   50  &     6.446 \\ 
15  &     3.083  &   33  &     5.483  &   51  &     6.464 \\ 
16  &     3.253  &   34  &     5.574  &   52  &     6.477 \\ 
17  &     3.419  &   35  &     5.662  &   53  &     6.485 \\ 
\hline
\end{tabular}
\caption{Energies of the vibrational levels of the electronic ground state $\mathrm{X}\,^2\Pi$ of the NO molecule for rotational quantum number $j=0$. The dissociation energy for the ground vibrational level is $D_0=6.490$~eV.  \label{tab:NOviblev}}
\end{table}

\section{Cross sections and rate constants \label{sec:results}}

\begin{figure}
\centering
\includegraphics[scale=.6]{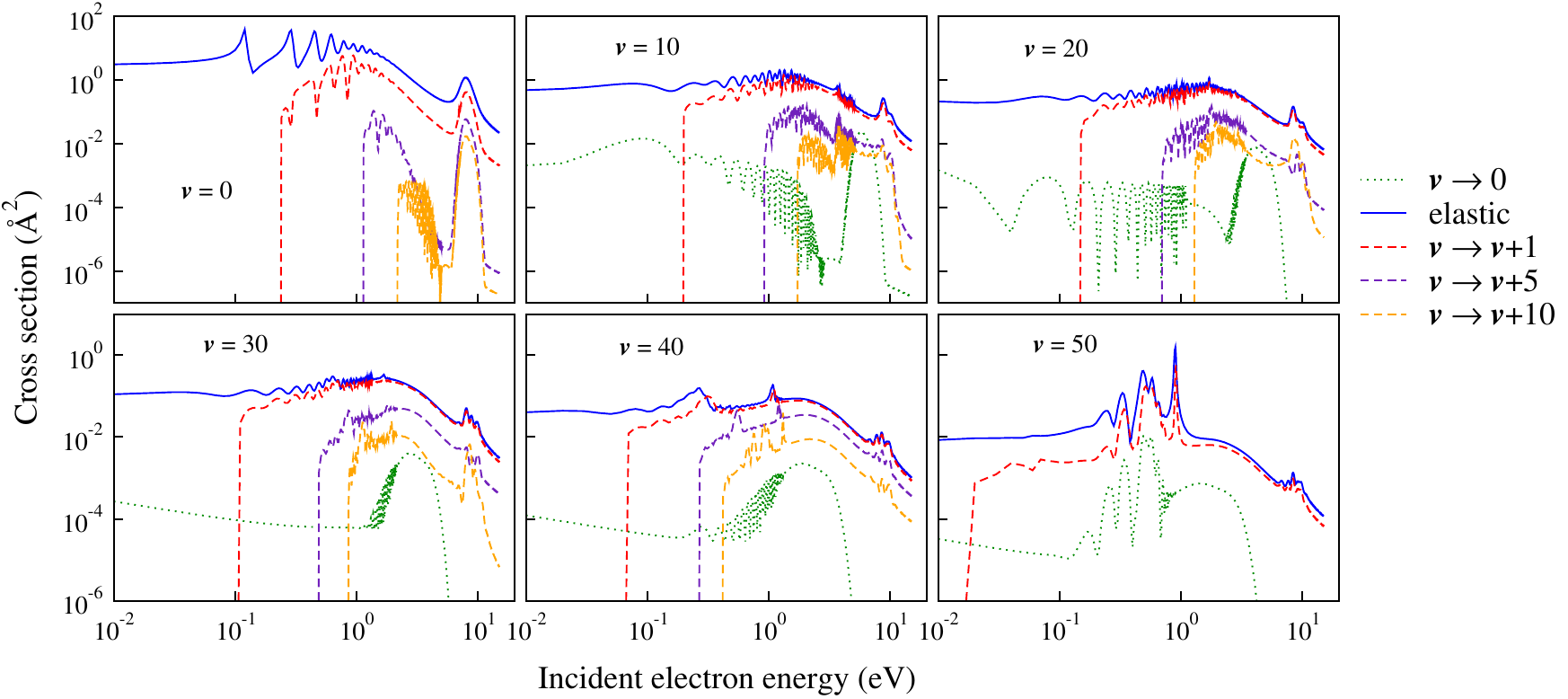} 
\caption{Summary on vibrational-excitation (VE dashed lines) and -de-excitation (VdE dotted lines) cross sections for NO molecule by electron impact as obtained in LCP model. Solid lines refer to the elastic processes. \label{fig:allxsec}}
\end{figure}

By using the molecular data presented in the Section~\ref{sec:th}, we report in Figure~\ref{fig:allxsec} a sample of the full set of vibrational-excitation (VE) and and de-excitation (VdE) cross sections calculated in the LCP approach, for the initial vibrational levels of NO molecule shown in the panel as a function of the incident electron energy. As it happens in the case of other molecules and molecular ions (for sake of comparison, see for example the papers in Refs.~\cite{Laporta_2021, 0741-3335-58-1-014024, 0963-0252-25-6-06LT02}), the on-the-threshold low-energy behaviour of the VE and VdE cross sections is characterized by a series of peaks which are related to the formation of the vibrationally excited NO$^-$ resonant states. We note that, due to the mutual arrangement of the NO and NO$^-$ potentials, the resonant peaks are represented by very narrow spikes similarly to the case of molecular oxygen  \cite{0963-0252-22-2-025001}. This correspond to a clear signature for long lived NO$^-$ anion formation. Beyond the dissociating threshold, a huge structure around 10 eV, particularly evident for $v=0$, is present due to the $^3\Pi$ resonant state.

In order to verify the results, we confirmed that the VE processes (also known as inelastic processes) and the corresponding VdE transitions (also known as super-elastic transitions) were correctly related by the detailed balance principle by the following formula:
\begin{equation}
\sigma^{\mathrm{VdE}}_{v' \to v}(\epsilon') = \sigma^{\mathrm{VE}}_{v\to v'}(\epsilon'+\epsilon^{th}_{v\to v'})\,\frac{\epsilon'+\epsilon^{th}_{v\to v'}}{\epsilon'}\,,
\end{equation}
where $\epsilon^{th}_{v\to v'} = \epsilon_{v'} - \epsilon_v$ is the threshold for the VE process considered.

\begin{figure}
\centering
\includegraphics[scale=.4]{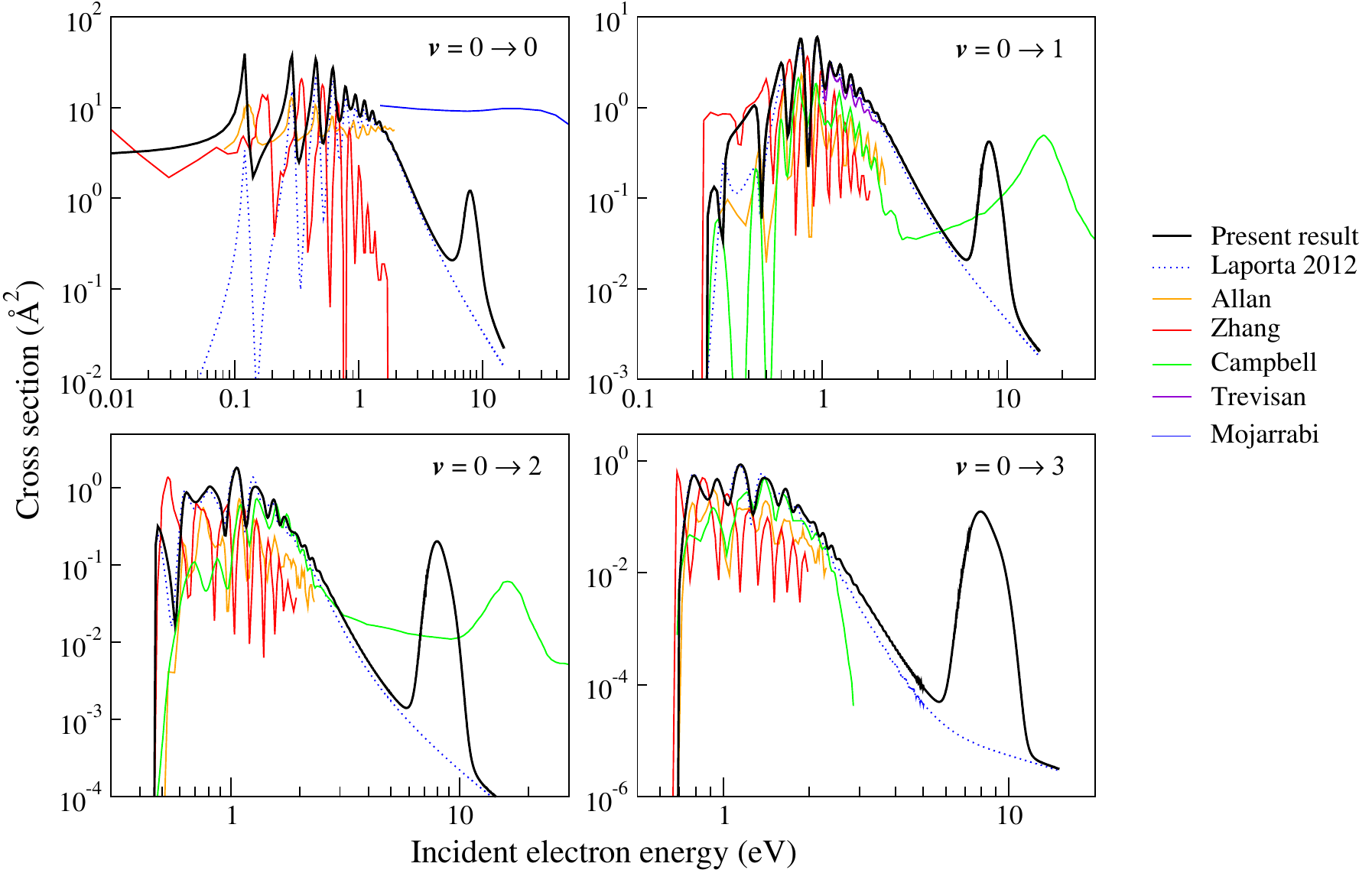}
\caption{Comparison of the cross sections obtained in the present paper (thick lines) with the data of Laporta-2012~\cite{0963-0252-21-5-055018}, Allan~\cite{0953-4075_38_5_011}, Zhang~\cite{PhysRevA.69.062711}, Campbell~\cite{doi:10.1029/2003GL019151}, Trevisan~\cite{PhysRevA.71.052714}  and Mojarrabi~\cite{0953-4075-28-3-019, doi:10.1063/1.4961372}  (thin lines) for the vibrational transitions shown in the panels.  \label{fig:xsec_comp}}
\end{figure}

To validate our calculations, in Figure~\ref{fig:xsec_comp} we give a comparison of our data with cross sections presented in the literature, for the  transitions shown in each panel. At low energies ($\lesssim 2$ eV), we remark a global good agreement with the experimental data of Allan~\cite{0953-4075_38_5_011} and Zhang~\cite{PhysRevA.69.062711}, and with the theoretical estimations of Trevisan~\cite{PhysRevA.71.052714} shown in solid thin lines. However, at high energies ($\gtrsim 2$ eV) our calculations show a single resonant peak around 10 eV, due to the interference of $^1\Pi$ and $^3\Pi$ symmetries of NO$^-$, and the comparison with the data of Mojarrabi~\cite{0953-4075-28-3-019, doi:10.1063/1.4961372} and Campbell~\cite{doi:10.1029/2003GL019151} reveals noticeable differences. In fact, as discussed in the experimental papers Refs.~\cite{doi:10.1063/1.447371, Szmytkowski_1991, 0953-4075-28-3-019}, the measurements show many and broad resonances in the region 10--20 eV and it is very difficult to link the observed resonant structures with the NO$^-$ states involved in the collision dynamics. For these reasons, we conclude that we would need to consider more resonant states above 10 eV in order to reproduce correctly the high energy behaviour of the cross sections.

Concerning the comparison with the previous LCP results obtained in the paper in Ref.~\cite{0963-0252-21-5-055018} (dotted lines in the graphs in Figure~\ref{fig:xsec_comp}), we note two main differences: (i) at low energies in the case of the $0\to0$ transition; and (ii) around 10 eV for all cases. The 10 eV structure is explicable since in the present work we take into account high lying resonant NO$^-$ states not present in Ref.~\cite{0963-0252-21-5-055018}. At the low energies, the discrepancy is due to the lack of the penetration factor $f_r$ in the present calculations, as explained in section~\ref{sec:th}. We note that the factor $f_r$ affects only the low energies and otherwise the cross sections are identical. Moreover, the lack of the \textit{ad hoc} factor $f_r$ in the $0\to0$ cross section improves the agreement with experimental results of Allan (orange line) and Zhang (red curve) at low energies and for this reason we decided to suppress that factor.

\begin{figure}
\centering
\includegraphics[scale=.55]{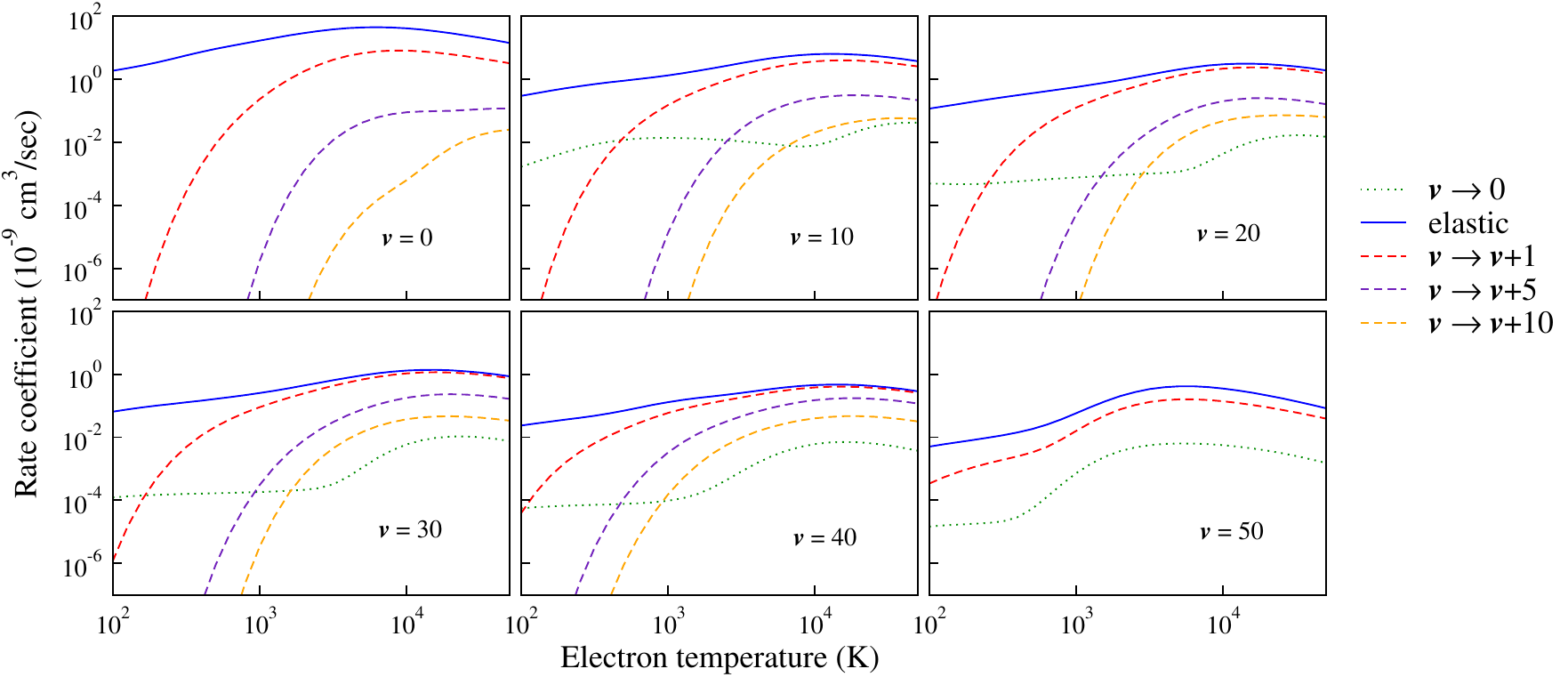} 
\caption{Summary on vibrational-excitation rate coefficients for electron-NO collisions. \label{fig:allrate}}
\end{figure}

In Figure~\ref{fig:allrate} we report the vibrationally resolved rate coefficients for electron-NO scattering as a function of the electron temperature calculated assuming a Maxwellian distribution for the electron energy and making a convolution with the cross sections in Figure~\ref{fig:allxsec}. We can verify that the rate coefficients decrease as the jump in quantum number is increases, as it would be expected. Nevertheless, for electron temperatures of the order of the eV, typical for gas discharges, the rate coefficients associated with jumps up to 10 quanta remain significant.

\begin{figure}
\centering
\includegraphics[scale=.4]{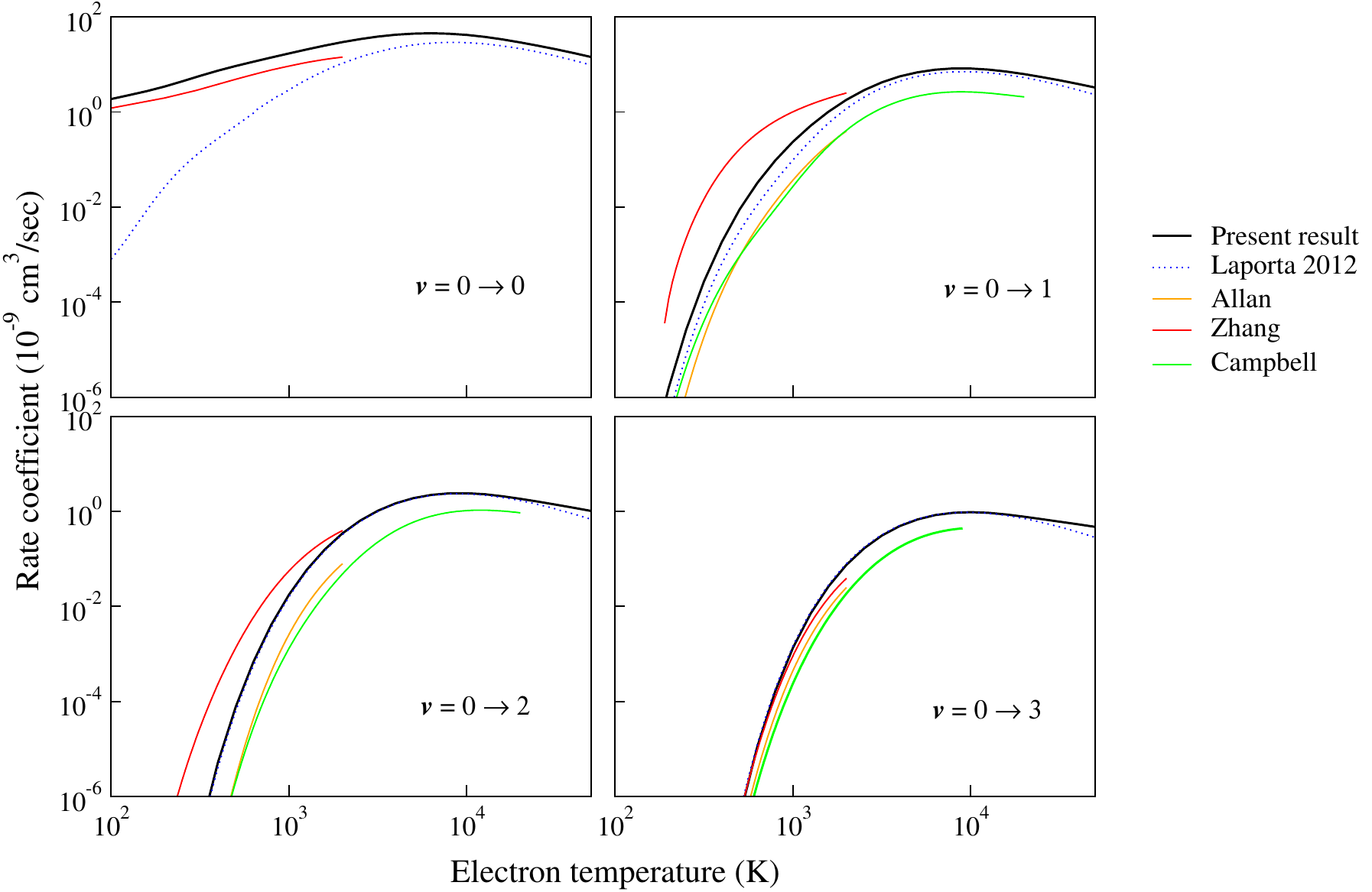}
\caption{Comparison of the rate coefficients obtained in the present paper (thick lines) with the data of Laporta-2012~\cite{0963-0252-21-5-055018}, Allan~\cite{0953-4075_38_5_011}, Zhang~\cite{PhysRevA.69.062711} and Campbell~\cite{doi:10.1029/2003GL019151} (thin lines) for the vibrational transitions shown in the panels.  \label{fig:rate_comp}}
\end{figure}

Figure \ref{fig:rate_comp} shows a comparison between rate constants calculated in the present work with those obtained from some of the experimental cross sections reported in Figure \ref{fig:xsec_comp}. Concerning the elastic transition, there is a significant discrepancy with the previous LCP calculations (dotted line)~\cite{0963-0252-21-5-055018} -- more pronounced for the lower values of the electron temperature, due to the penetration factor -- but it can be further confirmed we now obtain a better agreement with the experimental data of Zhang~\cite{PhysRevA.69.062711}. For the other inelastic transitions, the small difference with Laporta-2012~\cite{0963-0252-21-5-055018} is due to the presence of the resonant peak at 10 eV in the new calculations and in general the agreement with experimental data is quite good.

\section{Electron-vibration energy exchange \label{sec:relaxvib}}

In this section, we apply the LCP results for the electron-NO collisions presented in Section~\ref{sec:results} to investigate the \textit{electron-vibration} (e-V) energy exchanges, using  a State-to-State (StS) kinetics formalism in chemical non-equilibrium systems~\cite{doi:10.1063/1.1700221, B715095D}.
We consider a gas containing free-electrons and NO molecules making the following assumptions: (i) dissociation of NO can be neglected and only vibrational excitations are allowed, (ii) free-electrons are described by a Maxwellian distribution at fixed temperature $T_e$, (iii) the whole system is kept in a isothermal bath at constant free-electron temperature $T_e$, and (iv) at beginning the population of vibrational levels of NO molecule follows a Boltzmann distribution at temperature $T_{vib}^0 \neq T_e$.  We study the temporal evolution of the vibrational distribution function (VDF) of NO molecules from the initial Boltzmann distribution $T_{vib}^0$ toward the equilibrium at temperature $T_e$, as a result of the VE processes and hence no vibrational energy transfers in collisions involving two NO molecules, the so-called V-T (vibration-translation) and V-V (vibration-vibration) energy exchanges, are considered. Recently, a similar StS kinetics approach was applied to determine the e-V relaxation for systems containing N$_2$~\cite{laporta:104319, doi:10.1063/1.4900508} and O$_2$~\cite{Alves2016, Laporta201644} molecules and for V-V and V-T relaxations for NO molecules~\cite{doi:10.2514/1.T6462, doi:10.1063/1.1730368}.

In a zero-dimensional (0D)  StS-model, the time-dependence of the VDF, $n_{v}(t)$,  of NO molecules is obtained by solving the following set of master equations~\cite{doi:10.1063/1.1700221, B715095D}:
\begin{equation}
\frac{d n_{v}}{d t} = n_e \sum_{w=0}^{53} \left[k_{w,v}\,n_{w} - k_{v,w}\,n_{v}\right]\,, \qquad v=0,\ldots 53\,, \label{eq:master_eq}
\end{equation}
where the $k_{v,w}$ are the VE and VdE rate coefficients shown in Figure~\ref{fig:allrate} and $n_e$ is the electron density.

\begin{figure}
\centering
\includegraphics[scale=.35]{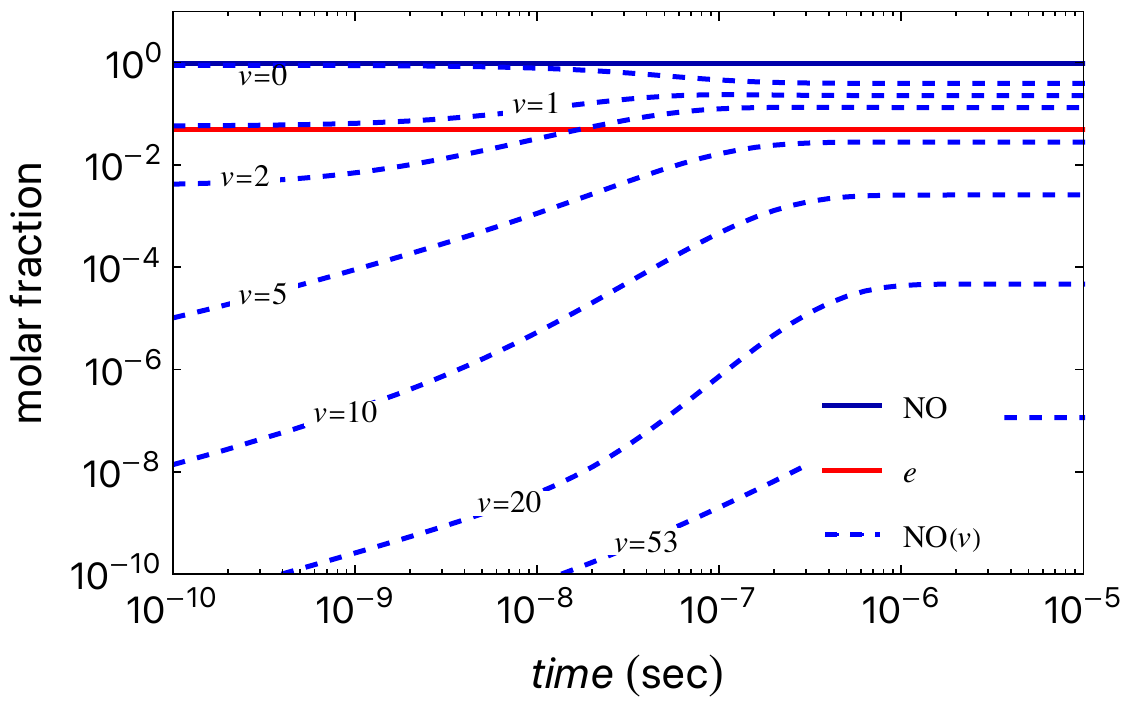} \includegraphics[scale=.34]{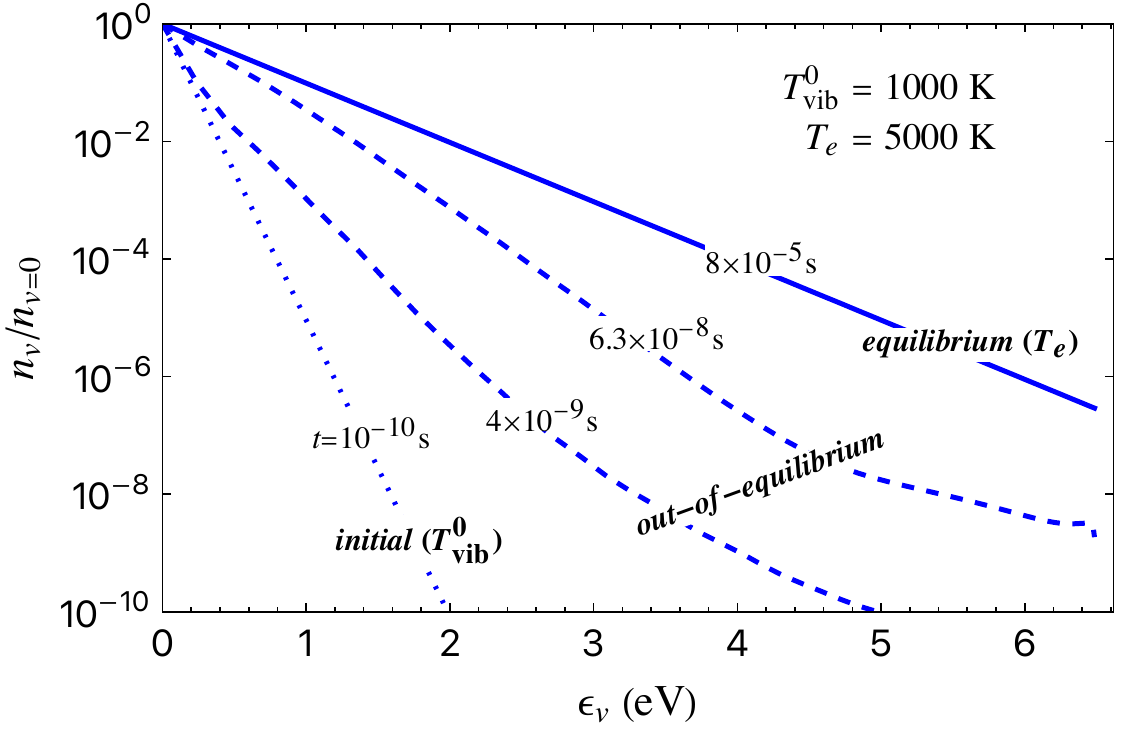} 
\\
\includegraphics[scale=.35]{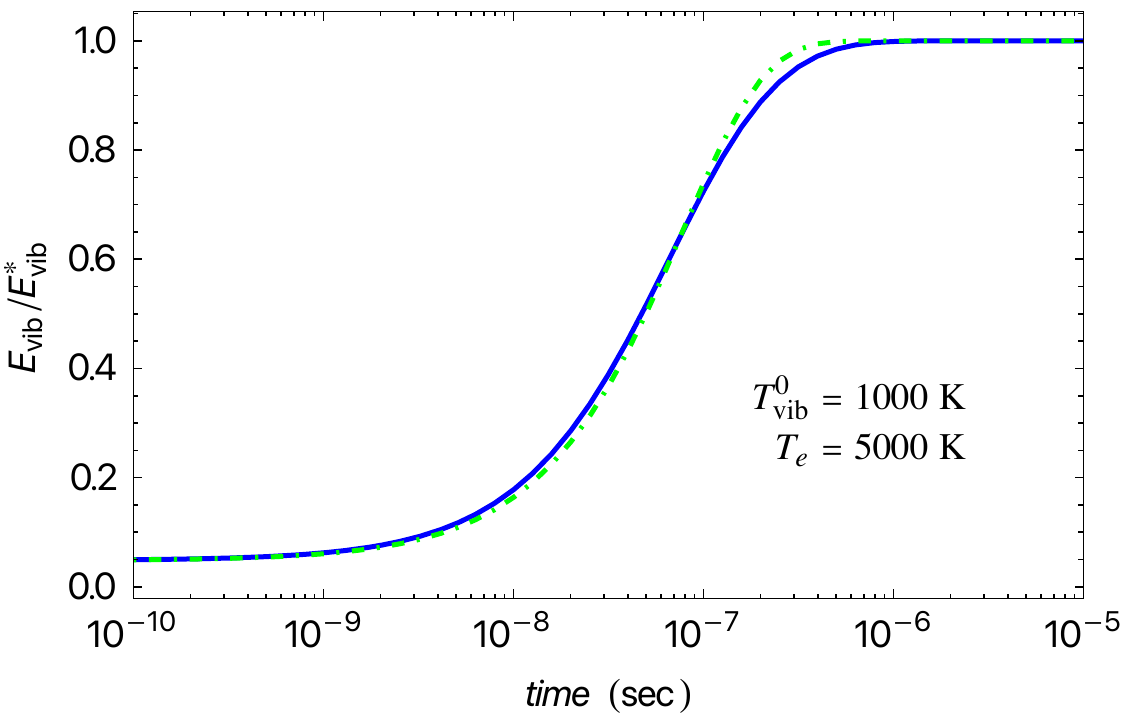} 
\caption{Electron-vibration energy exchange for a NO plasma out of equilibrium. (plot on the top-left) Time evolution of State-to-State molar fractions and (plot on the top-right) non-equilibrium vibrational distribution function of NO molecules. (plot on the bottom) Time evolution of the vibrational energy (solid line) compared with Landau-Teller approximation (broken line). The simulation corresponds to the electron temperature of $T_e=5000$~K and to the initial NO vibrational temperature of $T_{vib}^0=1000$~K.  \label{fig:timenoneq}}
\end{figure}

On the top of Figure~\ref{fig:timenoneq}, we show the time evolution for the molar fractions for the case $T_{vib}^0=1000$~K and $T_e=5000$~K,  for both NO-vibrationally-specific (dashed lines) and total-NO (solid line), and for the non-equilibrium VDF. In the calculations we set a total particle density of $n_0=3.2\times 10^{22}$ m$^{-3}$  with a concentration of electrons $n_e= 0.05\, n_0$. The time evolution of the VDF can be seen in the plot on the right in Figure~\ref{fig:timenoneq}, as the system starts from a Boltzmann equilibrium distribution at 1000 K (dotted line) and it ends onto a Boltzmann distribution at 5000 K (solid line) attained already at time $t = 8 \times 10^{-5}$ s. In the transient (dashed lines), strong non-equilibrium effects can be noted for the vibrational distribution and, in particular, the high vibrational levels are over populated. This behaviour can be explicitly noted in the time evolution of the molar fractions of vibrational level of NO, plot on the left of Figure~\ref{fig:timenoneq}: in fact the first vibrational levels, $v<5$, reach the equilibrium faster with respect to the other ones. 

In order to determine the e-V relaxation time for the NO molecule we assume a first order response of the system, where the characteristic relaxation time $\tau_e$ is defined similarly to the Landau-Teller (LT) approach for V-T energy exchanges~\cite{landauteller}, by:
\begin{equation}
\frac{dE_{vib}}{dt} = \frac{E^*_{vib}-E_{vib}}{\tau_e}\,,\label{eq:landauteller}
\end{equation}
where $E_{vib}$ is the vibrational energy of the NO molecules at time $t$:
\begin{equation}
E_{vib}(t) = \sum_{v=0}^{53}  n_{v}(t)\,\epsilon_{v}\,,\label{eq:Evib}
\end{equation}
$E_{vib}^*$ is the (final) equilibrium vibrational energy at temperature $T_e$:
\begin{equation}
E^*_{vib} = \sum_{v=0}^{53} n^*_{v}(T_e)\,\epsilon_{v}\,,\label{eq:Evibstar}
\end{equation}
and $n^*_{v}(T_e)$ is the equilibrium VDF of NO molecule.

In the plot on the bottom of Figure~\ref{fig:timenoneq} we report the vibrational energy of NO molecules as a function of the time (solid line), for the case considered above, $T_{vib}^0=1000$~K and $T_e=5000$~K, compared with LT approximation (broken curve). We remark as the linear approximation works very well, and it gives for the e-V relaxation time of NO a value of $\tau_e=7.7\times10^{-8}$ sec.

\begin{figure}
\centering
\includegraphics[scale=.9]{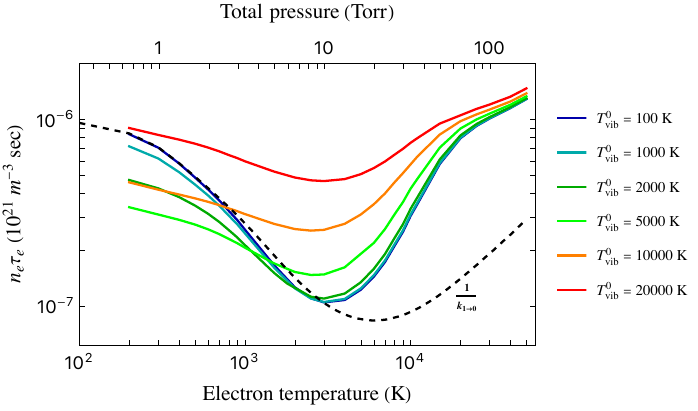} 
\caption{Electron-vibration relaxation time for NO molecule as obtained in the StS approach as a function of the electron temperature, for different  values of the initial vibrational temperature $T_{vib}^0$. For comparison, the inverse of the rate coefficient $k_{1\to 0}$ is reported (dashed line). \label{fig:relaxtime}}
\end{figure}

We calculated the e-V relaxation time of NO in LT approach for different configurations. In particular, we consider electron temperatures in the range $200<T_e<50000$ K, which corresponds to a total pressure between $0.66 - 165$ Torr, and the initial vibrational temperature in the interval $100<T^0_{vib}<20000$ K. The plot in Figure~\ref{fig:relaxtime} summarizes our results (solid curves). In order to make the e-V relaxation time independent to the electron density $n_e$,  we plot the product $n_e\tau_e$ as a functions of the electron temperature $T_e$, for different values of the initial vibrational temperature $T_{vib}^0$. The $n_e\tau_e$ product can be interpreted as an inverse ``effective'' rate coefficient integrated over all the vibrational levels. It can be noted from that the curves corresponding to $T_{vib}^0\lesssim 2000$ K are basically independent to $T_{vib}^0$ over all values of the electron temperature; moreover, for $T_e<3000$ K, the rate of e-V relaxation $n_e\tau_e$ is well described by the inverse of the fundamental deexcitation rate $k_{1\to0}$ (dashed curve) as predicted by Landau~\cite{doi:10.1063/1.1700221}. On the opposite range, for $T_e>20000$ K,  all curves of $T_{vib}^0$ asymptotically collapse into a single curve. For all other values of $T_e$ and $T_{vib}^0$ the response of the system is no longer linear and the relaxation is not well described by a single-value function. However, the plot in Figure~\ref{fig:relaxtime} clearly indicates a collective behaviour of the vibrational levels and that the full set of the vibrational excitation rate coefficients can be reduced in order to describe the transport phenomena and kinetics of more complex systems. In order to give an analytic form for the relaxation time, in Table \ref{tab:fitrelax} we present a fit for the curves in Figure~\ref{fig:relaxtime}.

From our calculations, and for the temperature interval considered, we find an e-V relaxation time in the range $6.5\times 10^{-8}\lesssim \tau_e \lesssim 10^{-6}$ sec. The only experimental data available in literature that we have found for the electron-NO relaxation time are the measurements by Narasinga \textit{et al.}~\cite{NARASINGARAO1968296}, albeit referring to very low pressures ($<1$ Torr). Considering a pressure of 0.66 Torr ($T_e=200$ K), we obtain $\tau_e=0.5\times 10^{-6}$ sec for $T_{vib}^0=100$ K which is compatible with the results of $0.3\times 10^{-6}$ sec reported in the Figure 1 of Narasinga paper. We observe that the difference in value can arise from the NO molecular rotational effects which become important at low pressure and temperature range. Concerning the energy exchanges between heavy particle, for NO-NO and NO-O systems, we note that the vibrational relaxation time for NO molecule by electron collisions is of the same order of magnitude of $\tau_v$ -- for V-V and V-T relaxations -- as reported in the papers in Refs.~\cite{doi:10.2514/1.T6462, doi:10.1063/1.1730368}.

\begin{table}
\centering
\begin{tabular}{ccccc}
\hline
~~$T_{vib}^0$~~ & ~~$a_0\times 10^{-6}$~~ & ~~$a_1 \times 10^{-11}$~~ & ~~$a_2\times 10^{-16}$~~ & ~~$b \times 10^{-5}$~~\\
\hline
100     & $-2.05$ & $7.68$ & $-7.87$ & $1.54$ \\
1000   & $-1.82$ & $7.31$ & $-7.38$ & $1.37$ \\
2000   & $-1.31$ & $6.58$ & $-6.45$ & $0.99$ \\
5000   & $-0.83$ & $6.22$ & $-6.32$ & $0.65$ \\
10000 & $-0.64$ & $6.06$ & $-6.44$ & $0.60$ \\
20000 & $-0.78$ & $5.54$ & $-5.65$ & $0.91$ \\
\hline
\end{tabular}
\caption{Fit  $a_0+a_1\,T_e + a_2\,T_e^2 + b\,/\log (T_e)$ for the curves reported in Figure~\ref{fig:relaxtime}. \label{tab:fitrelax}}
\end{table}

\section{Non-equilibrium EEDFs \label{sec:EEDF}}

Electron-impact cross sections are needed for the numerical solution of the electron Boltzmann equation,  as to obtain the non-equilibrium Electron Energy Distribution Functions (EEDFs) and related macroscopic quantities, such as the electron swarm parameters~\cite{hagelaar2005solving}.  In this Section, we calculate non-equilibrium EEDFs and swarm parameters for electrons in a pure NO system, by using three different sets of cross sections. These three sets differ only for the electron-impact excitation and de-excitation cross sections for the first four vibrational excitations of the ground state (\textit{i.e.} NO$(\mathrm{X}, v=0) \rightarrow $ NO$(\mathrm{X}, v')$, with $v' = 1 - 3$): (i) the first set uses the cross sections presented in the present work in Section~\ref{sec:results}, (ii) the second set includes electron-impact excitation cross sections by Zhang and co-authors \cite{PhysRevA.69.062711} and (iii) the third one by Campbell and co-authors \cite{doi:10.1029/2003GL019151}. Differences in shape and magnitude of the aforementioned cross sections are shown in Figure~\ref{fig:xsec_comp}. We point out that, for the swarm conditions under considerations, excitations of higher vibrational levels (\textit{i.e.} NO$(\mathrm{X}, v' \geq 4)$) do not influence the calculations of electron swarm parameters by more than $1\%$ and they are neglected in the present calculations. Nevertheless, high vibrational levels excitations are important in plasma modelling when considering conditions of high vibrational-translational non-equilibrium~\cite{BOUZIANE2022905}. For the other processes the cross sections data are taken from the Hayashi database of LXCat~\cite{Hayashi}, that includes two electronic excitation cross sections, one total ionization cross section, and one electron attachment cross section leading to the formation of NO$^-$. The effective cross section from the Hayashi database has been replaced with an elastic momentum transfer cross section and 122 rotational excitations and de-excitations cross sections based on the Born approximation, as described by Song and co-authors~\cite{doi:10.1063/1.5114722}.

In the calculations, a Boltzmann population at 300 K for the vibrational and the rotational levels of the electronic ground state of NO molecule has been considered and cross sections for superelastic collisions from rotational and vibrational levels have been calculated using the Klein-Rosseland formula~\cite{fowler1924xxiii}.  Additionally,  anisotropic scattering for rotational collisions has been considered by taking into account rotational integral and momentum transfer cross sections, as described by Vialetto and co-authors~\cite{vialetto2021effect}. The use of an elastic momentum transfer cross section, instead of an effective one, is preferred for this Section, as the only difference between results obtained from the three sets comes from the choice of vibrational excitations and de-excitations cross sections.

The electron Boltzmann equation is solved numerically using the Lisbon KInetics two-term Boltzmann solver (LoKI-B)~\cite{tejero2019lisbon} considering stationary conditions and a constant reduced electric field ($E/N$). All the calculations are performed assuming a temporal growth of the electron density and adopting 1000 energy cells that are automatically adjusted by prescribing the decade-fall of the electron energy distribution function between 20 and 25.

\begin{figure}
\centering
\includegraphics[scale=.9]{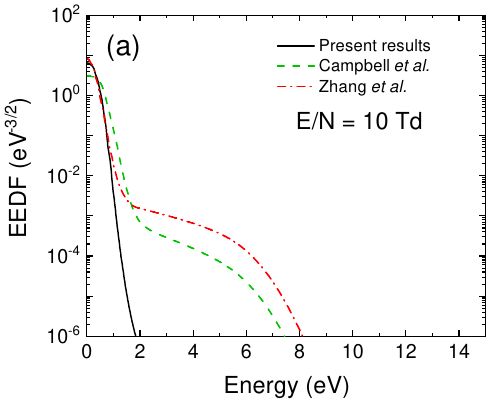}
\includegraphics[scale=.9]{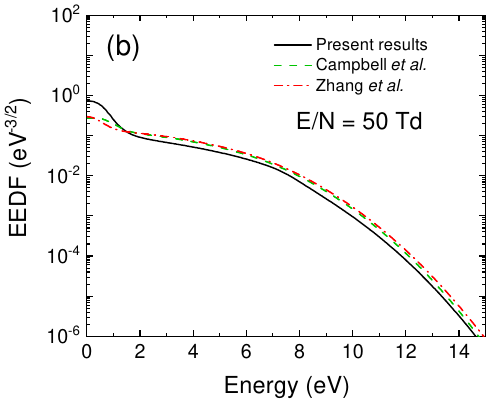}
\caption{Calculated EEDFs with LoKI-B solver \cite{tejero2019lisbon} at (a) $E/N = 10$ Td and (b) $E/N = 50$ Td. Three sets of vibrational excitation cross sections have been used in the calculations, including the ones presented in this work (solid line), the ones by Campbell and co-authors~\cite{doi:10.1029/2003GL019151} (dashed line), and by Zhang and co-authors~\cite{PhysRevA.69.062711} (dot-dashed line).}
\label{fig:eedf}
\end{figure}

The calculated EEDFs for $E/N = 10$ Td and $50$ Td are shown in the Figures~\ref{fig:eedf}(a) and (b), respectively.  For both $E/N$ values, large differences are found between the EEDFs calculated using the cross sections from the present work and the ones from Campbell and co-authors~\cite{doi:10.1029/2003GL019151} or Zhang and co-authors~\cite{PhysRevA.69.062711}. These discrepancies are more pronounced at 10 Td (Figure~\ref{fig:eedf}(a)), as vibrational excitations and de-excitations are the dominant energy exchanges for this value of $E/N$.  At 50 Td (Figure~\ref{fig:eedf}(b)), a long tail of the EEDF is present due to the higher value of $E/N$. Moreover, the magnitude of the bulk of the EEDF (\textit{i.e.} below 2 eV) obtained with the present cross sections is higher than the one calculated when using cross sections from the other two sources. This feature is due to the fact that the present cross sections for vibrational excitation have typically higher magnitude than the ones by Zhang and co-authors~\cite{PhysRevA.69.062711} and by Campbell and co-authors~\cite{doi:10.1029/2003GL019151} (see Figure~\ref{fig:xsec_comp}). As such,  those inelastic processes act as a sharp barrier that electrons have to overcome to reach higher energies.

\begin{figure}
\centering
\includegraphics[scale=.9]{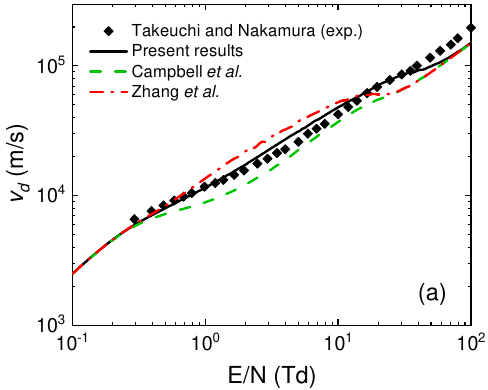}
\includegraphics[scale=.9]{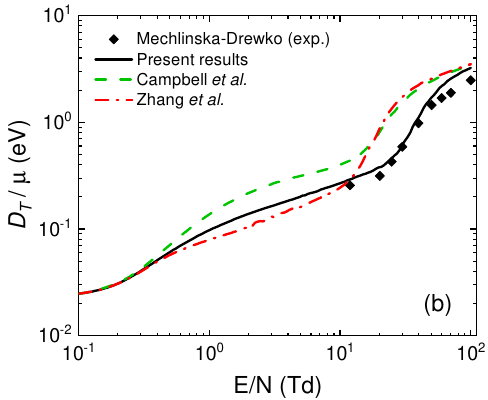}
\caption{(a) Electron drift velocity and (b) characteristic energy for a NO gas at 300 K, as a function of the reduced electric field. Calculations have been performed using the LoKI-B solver~\cite{tejero2019lisbon} for the three set of cross sections: by the present work (solid line), by Campbell and co-authors~\cite{doi:10.1029/2003GL019151} (dashed line) and by Zhang and co-authors~\cite{PhysRevA.69.062711} (dot-dashed line). Experimental results (diamonds) are taken from the papers of Takeuchi~\cite{takeuchi2001measurements} and Mechlinska~\cite{mechlinska1999dt}.} 
\label{fig:swarm_parameters}
\end{figure}

The calculated electron drift velocity ($v_d$) and characteristic energy ($D_T/\mu$) in a pure NO gas at 300 K are plotted in Figures~\ref{fig:swarm_parameters}(a) and \ref{fig:swarm_parameters}(b), respectively, for 0.1 Td $\leq E/N \leq 100$ Td.  A comparison with measurements of Takeuchi and Nakamura \cite{takeuchi2001measurements} and Mechlinska-Drewko and co-authors \cite{mechlinska1999dt} is also shown. Concerning the electron drift velocity (Figure~\ref{fig:swarm_parameters}(a)), all the three cross section sets reproduce the general trend of the measurements. However, between 0.3 Td and 80 Td, the magnitude and shape of the vibrational excitation cross sections influence the value of the calculated $v_d$, leading to deviations up to $32\%$ between results obtained with the present cross sections and the ones from Campbell and co-authors~\cite{doi:10.1029/2003GL019151}. For the electron characteristic energy (Figure~\ref{fig:swarm_parameters}(b)), we should point out discrepancies, up to a factor 3, between the results of $D_T/\mu$ obtained using Campbell \textit{et al.}~\cite{doi:10.1029/2003GL019151} or Zhang \textit{et al.}~\cite{PhysRevA.69.062711} with respect to experimental measurements~\cite{mechlinska1999dt}. In turn, using the cross sections proposed in this work leads to a much better agreement with the experimental data. This is due to the lower magnitude of the vibrational excitation cross sections from Ref.~\cite{doi:10.1029/2003GL019151,PhysRevA.69.062711} compared with the ones presented in this work. 

The results in Figure~\ref{fig:swarm_parameters} evince the improvement in calculated data brought by the cross sections for process (\ref{eq:VEprocess}) proposed here. Nevertheless, it is important to note that the electron impact cross sections used in this work do not constitute a consistent set, as the ones derived from swarm analysis~\cite{petrovic2009measurement}. Examples of other cross sections sets derived from the swarm analysis technique can be found in the work by Song and co-authors~\cite{doi:10.1063/1.5114722}, Stokes and co-authors~\cite{stokes2021toward}, and Josi{\'c} and co-authors~\cite{Josic2001318}. Therefore, this work highlights the importance of having accurate electron impact cross sections for vibrational excitations of the NO molecule, their impact in the calculation of non-equilibrium EEDFs and electron swarm parameters, and strongly suggest the validity of our calculations. Further work should focus on refining the cross section set used here to define a complete and consistent cross section set that includes the present VE cross sections.

\section{Conclusions \label{sec:conc}}

In this work, we calculated a new set of \textit{ab initio} electron impact cross sections and rate coefficients for the vibrational excitation (VE) process of NO molecule for its ground electronic state (X\,$^2\Pi$). These cross sections were used to study electron-vibration (e-V) energy exchanges in non-equilibrium thermodynamics conditions. In particular, a state-to-state model describing electron impact excitation and de-excitaiton of the individual vibrational levels NO(X\,$^2\Pi, v$) was used to derive macroscopic quantities as the vibrational relaxation time. The model predictions are compatible with the available experimental data, providing a first validation of our calculations. Moreover, we find a vibrational relaxation time due to electrons -- \textit{i.e.} the e-V process -- comparable with the relaxation in V-V and V-T energy exchanges due to heavy-particle collisions.

Our vibrational excitation cross sections were incorporated on a default set of cross sections to calculate electron swarm parameters, comparing the characteristic energy and the electron drift velocity in NO with existing experimental measurements. The results indicate that our cross sections, when used within the Hayashi set of electron impact cross sections, significantly improve the agreement between the calculated and measured data, reinforcing the validity of the VE cross sections proposed here. Further investigation is needed to derive a complete and consistent set of electron impact cross section for NO that includes the vibrational excitation cross sections presented in this work. Nevertheless, the present results indicate that the use of our VE cross sections, combined with electron impact cross sections from the Hayashi database of LXCat~\cite{Hayashi}, already form a good starting point to develop a complete and consistent set.

The full set of data presented in this work are available on LXCat database~\cite{LxCat_Laporta}.

\section*{Acknowledgements}
VG was partially supported by the Portuguese FCT - Funda\c{c}\~{a}o para a Ci\^{e}ncia e a Tecnologia, under Projects UIDB/50010/2020, UIDP/50010/2020 and PTDC/FIS-PLA/1616/2021 (PARADiSE). VL thanks Prof. Ioan Schneider (CNRS--Universit{\'{e}} Le Havre, France) for careful reading of the manuscript.


\begin{thebibliography}{51}
\expandafter\ifx\csname natexlab\endcsname\relax\def\natexlab#1{#1}\fi
\expandafter\ifx\csname bibnamefont\endcsname\relax
  \def\bibnamefont#1{#1}\fi
\expandafter\ifx\csname bibfnamefont\endcsname\relax
  \def\bibfnamefont#1{#1}\fi
\expandafter\ifx\csname citenamefont\endcsname\relax
  \def\citenamefont#1{#1}\fi
\expandafter\ifx\csname url\endcsname\relax
  \def\url#1{\texttt{#1}}\fi
\expandafter\ifx\csname urlprefix\endcsname\relax\def\urlprefix{URL }\fi
\providecommand{\bibinfo}[2]{#2}
\providecommand{\eprint}[2][]{\url{#2}}



\bibitem[{\citenamefont{Bouziane et~al.}(2022)\citenamefont{Bouziane, Ferdi,
  and Djebli}}]{BOUZIANE2022905}
\bibinfo{author}{\bibfnamefont{A.}~\bibnamefont{Bouziane}},
  \bibinfo{author}{\bibfnamefont{M.~A.} \bibnamefont{Ferdi}}, \bibnamefont{and}
  \bibinfo{author}{\bibfnamefont{M.}~\bibnamefont{Djebli}},
  \bibinfo{journal}{Advances in Space Research} \textbf{\bibinfo{volume}{69}},
  \bibinfo{pages}{905} (\bibinfo{year}{2022}).




\bibitem[{\citenamefont{Sahai et~al.}(2020)\citenamefont{Sahai, Johnston,
  Lopez, and Panesi}}]{SAHAI2020106752}
\bibinfo{author}{\bibfnamefont{A.}~\bibnamefont{Sahai}},
  \bibinfo{author}{\bibfnamefont{C.~O.} \bibnamefont{Johnston}},
  \bibinfo{author}{\bibfnamefont{B.}~\bibnamefont{Lopez}}, \bibnamefont{and}
  \bibinfo{author}{\bibfnamefont{M.}~\bibnamefont{Panesi}},
  \bibinfo{journal}{Journal of Quantitative Spectroscopy and Radiative
  Transfer} \textbf{\bibinfo{volume}{242}}, \bibinfo{pages}{106752}
  (\bibinfo{year}{2020}).

\bibitem[{\citenamefont{Munaf{\`o} et~al.}(2015)\citenamefont{Munaf{\`o},
  Alfuhaid, Cambier, and
  Panesi}}]{:/content/aip/journal/jap/118/13/10.1063/1.4931769}
\bibinfo{author}{\bibfnamefont{A.}~\bibnamefont{Munaf{\`o}}},
  \bibinfo{author}{\bibfnamefont{S.~A.} \bibnamefont{Alfuhaid}},
  \bibinfo{author}{\bibfnamefont{J.-L.} \bibnamefont{Cambier}},
  \bibnamefont{and} \bibinfo{author}{\bibfnamefont{M.}~\bibnamefont{Panesi}},
  \bibinfo{journal}{Journal of Applied Physics} \textbf{\bibinfo{volume}{118}},
  \bibinfo{eid}{133303} (\bibinfo{year}{2015}).

\bibitem[{\citenamefont{Panesi and Lani}(2013)}]{doi:10.1063/1.4804388}
\bibinfo{author}{\bibfnamefont{M.}~\bibnamefont{Panesi}} \bibnamefont{and}
  \bibinfo{author}{\bibfnamefont{A.}~\bibnamefont{Lani}},
  \bibinfo{journal}{Physics of Fluids} \textbf{\bibinfo{volume}{25}},
  \bibinfo{pages}{057101} (\bibinfo{year}{2013}).






\bibitem[{\citenamefont{Naseem}(2005)}]{NASEEM200533}
\bibinfo{author}{\bibfnamefont{K.~M.} \bibnamefont{Naseem}},
  \bibinfo{journal}{Molecular Aspects of Medicine}
  \textbf{\bibinfo{volume}{26}}, \bibinfo{pages}{33} (\bibinfo{year}{2005}).

\bibitem[{\citenamefont{Feng}(2012)}]{Feng:2012wq}
\bibinfo{author}{\bibfnamefont{C.}~\bibnamefont{Feng}},
  \bibinfo{journal}{Coordination chemistry reviews}
  \textbf{\bibinfo{volume}{256}}, \bibinfo{pages}{393} (\bibinfo{year}{2012}).

\bibitem[{\citenamefont{Moncada and
  Higgs}(2006)}]{https://doi.org/10.1038/sj.bjp.0706458}
\bibinfo{author}{\bibfnamefont{S.}~\bibnamefont{Moncada}} \bibnamefont{and}
  \bibinfo{author}{\bibfnamefont{E.~A.} \bibnamefont{Higgs}},
  \bibinfo{journal}{British Journal of Pharmacology}
  \textbf{\bibinfo{volume}{147}}, \bibinfo{pages}{S193} (\bibinfo{year}{2006}).

\bibitem[{\citenamefont{Jeon et~al.}(2014)\citenamefont{Jeon, Sharma, Seo, and
  Kwon}}]{Jeon:2014tu}
\bibinfo{author}{\bibfnamefont{K.-J.} \bibnamefont{Jeon}},
  \bibinfo{author}{\bibfnamefont{R.}~\bibnamefont{Sharma}},
  \bibinfo{author}{\bibfnamefont{J.-W.} \bibnamefont{Seo}}, \bibnamefont{and}
  \bibinfo{author}{\bibfnamefont{S.}~\bibnamefont{Kwon}},
  \bibinfo{journal}{Journal of Nanomaterials} \textbf{\bibinfo{volume}{2014}},
  \bibinfo{pages}{523646} (\bibinfo{year}{2014}).

\bibitem[{\citenamefont{Chen et~al.}(2021)\citenamefont{Chen, Yuan, Wu, Lin,
  and Li}}]{Chen:2021tk}
\bibinfo{author}{\bibfnamefont{H.}~\bibnamefont{Chen}},
  \bibinfo{author}{\bibfnamefont{D.}~\bibnamefont{Yuan}},
  \bibinfo{author}{\bibfnamefont{A.}~\bibnamefont{Wu}},
  \bibinfo{author}{\bibfnamefont{X.}~\bibnamefont{Lin}}, \bibnamefont{and}
  \bibinfo{author}{\bibfnamefont{X.}~\bibnamefont{Li}}, \bibinfo{journal}{Waste
  Disposal \& Sustainable Energy} \textbf{\bibinfo{volume}{3}},
  \bibinfo{pages}{201} (\bibinfo{year}{2021}).

\bibitem[{\citenamefont{Vervloessem et~al.}(2020)\citenamefont{Vervloessem,
  Aghaei, Jardali, Hafezkhiabani, and
  Bogaerts}}]{doi:10.1021/acssuschemeng.0c01815}
\bibinfo{author}{\bibfnamefont{E.}~\bibnamefont{Vervloessem}},
  \bibinfo{author}{\bibfnamefont{M.}~\bibnamefont{Aghaei}},
  \bibinfo{author}{\bibfnamefont{F.}~\bibnamefont{Jardali}},
  \bibinfo{author}{\bibfnamefont{N.}~\bibnamefont{Hafezkhiabani}},
  \bibnamefont{and} \bibinfo{author}{\bibfnamefont{A.}~\bibnamefont{Bogaerts}},
  \bibinfo{journal}{ACS Sustainable Chemistry \& Engineering}
  \textbf{\bibinfo{volume}{8}}, \bibinfo{pages}{9711} (\bibinfo{year}{2020}).

\bibitem[{\citenamefont{Rouwenhorst et~al.}(2021)\citenamefont{Rouwenhorst,
  Jardali, Bogaerts, and Lefferts}}]{D0EE03763J}
\bibinfo{author}{\bibfnamefont{K.~H.~R.} \bibnamefont{Rouwenhorst}},
  \bibinfo{author}{\bibfnamefont{F.}~\bibnamefont{Jardali}},
  \bibinfo{author}{\bibfnamefont{A.}~\bibnamefont{Bogaerts}}, \bibnamefont{and}
  \bibinfo{author}{\bibfnamefont{L.}~\bibnamefont{Lefferts}},
  \bibinfo{journal}{Energy Environ. Sci.} \textbf{\bibinfo{volume}{14}},
  \bibinfo{pages}{2520} (\bibinfo{year}{2021}).

\bibitem[{\citenamefont{Jardali et~al.}(2021)\citenamefont{Jardali, Van~Alphen,
  Creel, Ahmadi~Eshtehardi, Axelsson, Ingels, Snyders, and
  Bogaerts}}]{D0GC03521A}
\bibinfo{author}{\bibfnamefont{F.}~\bibnamefont{Jardali}},
  \bibinfo{author}{\bibfnamefont{S.}~\bibnamefont{Van~Alphen}},
  \bibinfo{author}{\bibfnamefont{J.}~\bibnamefont{Creel}},
  \bibinfo{author}{\bibfnamefont{H.}~\bibnamefont{Ahmadi~Eshtehardi}},
  \bibinfo{author}{\bibfnamefont{M.}~\bibnamefont{Axelsson}},
  \bibinfo{author}{\bibfnamefont{R.}~\bibnamefont{Ingels}},
  \bibinfo{author}{\bibfnamefont{R.}~\bibnamefont{Snyders}}, \bibnamefont{and}
  \bibinfo{author}{\bibfnamefont{A.}~\bibnamefont{Bogaerts}},
  \bibinfo{journal}{Green Chem.} \textbf{\bibinfo{volume}{23}},
  \bibinfo{pages}{1748} (\bibinfo{year}{2021}).

\bibitem[{\citenamefont{Bahnamiri et~al.}(2021)\citenamefont{Bahnamiri,
  Verheyen, Snyders, Bogaerts, and Britun}}]{Bahnamiri_2021}
\bibinfo{author}{\bibfnamefont{O.~S.} \bibnamefont{Bahnamiri}},
  \bibinfo{author}{\bibfnamefont{C.}~\bibnamefont{Verheyen}},
  \bibinfo{author}{\bibfnamefont{R.}~\bibnamefont{Snyders}},
  \bibinfo{author}{\bibfnamefont{A.}~\bibnamefont{Bogaerts}}, \bibnamefont{and}
  \bibinfo{author}{\bibfnamefont{N.}~\bibnamefont{Britun}},
  \bibinfo{journal}{Plasma Sources Science and Technology}
  \textbf{\bibinfo{volume}{30}}, \bibinfo{pages}{065007}
  (\bibinfo{year}{2021}).

\bibitem[{\citenamefont{Bak and Cappelli}(2015)}]{doi:10.1098/rsta.2014.0331}
\bibinfo{author}{\bibfnamefont{M.~S.} \bibnamefont{Bak}} \bibnamefont{and}
  \bibinfo{author}{\bibfnamefont{M.~A.} \bibnamefont{Cappelli}},
  \bibinfo{journal}{Philosophical Transactions of the Royal Society A:
  Mathematical, Physical and Engineering Sciences}
  \textbf{\bibinfo{volume}{373}}, \bibinfo{pages}{20140331}
  (\bibinfo{year}{2015}).


\bibitem[{\citenamefont{Campbell et~al.}(2004)\citenamefont{Campbell, Brunger,
  Petrovic, Jelisavcic, Panajotovic, and Buckman}}]{doi:10.1029/2003GL019151}
\bibinfo{author}{\bibfnamefont{L.}~\bibnamefont{Campbell}},
  \bibinfo{author}{\bibfnamefont{M.~J.} \bibnamefont{Brunger}},
  \bibinfo{author}{\bibfnamefont{Z.~L.} \bibnamefont{Petrovic}},
  \bibinfo{author}{\bibfnamefont{M.}~\bibnamefont{Jelisavcic}},
  \bibinfo{author}{\bibfnamefont{R.}~\bibnamefont{Panajotovic}},
  \bibnamefont{and} \bibinfo{author}{\bibfnamefont{S.~J.}
  \bibnamefont{Buckman}}, \bibinfo{journal}{Geophysical Research Letters}
  \textbf{\bibinfo{volume}{31}} (\bibinfo{year}{2004}).

\bibitem[{\citenamefont{Song et~al.}(2019)\citenamefont{Song, Yoon, Cho,
  Karwasz, Kokoouline, Nakamura, and Tennyson}}]{doi:10.1063/1.5114722}
\bibinfo{author}{\bibfnamefont{M.-Y.} \bibnamefont{Song}},
  \bibinfo{author}{\bibfnamefont{J.-S.} \bibnamefont{Yoon}},
  \bibinfo{author}{\bibfnamefont{H.}~\bibnamefont{Cho}},
  \bibinfo{author}{\bibfnamefont{G.~P.} \bibnamefont{Karwasz}},
  \bibinfo{author}{\bibfnamefont{V.}~\bibnamefont{Kokoouline}},
  \bibinfo{author}{\bibfnamefont{Y.}~\bibnamefont{Nakamura}}, \bibnamefont{and}
  \bibinfo{author}{\bibfnamefont{J.}~\bibnamefont{Tennyson}},
  \bibinfo{journal}{Journal of Physical and Chemical Reference Data}
  \textbf{\bibinfo{volume}{48}}, \bibinfo{pages}{043104}
  (\bibinfo{year}{2019}).

\bibitem[{\citenamefont{Itikawa}(2016)}]{doi:10.1063/1.4961372}
\bibinfo{author}{\bibfnamefont{Y.}~\bibnamefont{Itikawa}},
  \bibinfo{journal}{Journal of Physical and Chemical Reference Data}
  \textbf{\bibinfo{volume}{45}}, \bibinfo{pages}{033106}
  (\bibinfo{year}{2016}).

\bibitem[{\citenamefont{Armenise and Esposito}(2021)}]{ARMENISE2021111325}
\bibinfo{author}{\bibfnamefont{I.}~\bibnamefont{Armenise}} \bibnamefont{and}
  \bibinfo{author}{\bibfnamefont{F.}~\bibnamefont{Esposito}},
  \bibinfo{journal}{Chemical Physics} \textbf{\bibinfo{volume}{551}},
  \bibinfo{pages}{111325} (\bibinfo{year}{2021}).

\bibitem[{\citenamefont{Esposito and
  Armenise}(2021)}]{doi:10.1021/acs.jpca.0c09999}
\bibinfo{author}{\bibfnamefont{F.}~\bibnamefont{Esposito}} \bibnamefont{and}
  \bibinfo{author}{\bibfnamefont{I.}~\bibnamefont{Armenise}},
  \bibinfo{journal}{The Journal of Physical Chemistry A}
  \textbf{\bibinfo{volume}{125}}, \bibinfo{pages}{3953} (\bibinfo{year}{2021}).

\bibitem[{\citenamefont{Laporta
  et~al.}(2020{\natexlab{a}})\citenamefont{Laporta, Tennyson, and
  Schneider}}]{10.1088/1361-6595/ab86d8}
\bibinfo{author}{\bibfnamefont{V.}~\bibnamefont{Laporta}},
  \bibinfo{author}{\bibfnamefont{J.}~\bibnamefont{Tennyson}}, \bibnamefont{and}
  \bibinfo{author}{\bibfnamefont{I.~F.} \bibnamefont{Schneider}},
  \bibinfo{journal}{Plasma Sources Science and Technology}
  \textbf{\bibinfo{volume}{29}}, \bibinfo{pages}{05LT02}
  (\bibinfo{year}{2020}).
  
  
\bibitem[{\citenamefont{Laporta
  et~al.}(2020{\natexlab{b}})\citenamefont{Laporta, Schneider, and
  Tennyson}}]{Laporta_2020}
\bibinfo{author}{\bibfnamefont{V.}~\bibnamefont{Laporta}},
  \bibinfo{author}{\bibfnamefont{I.~F.} \bibnamefont{Schneider}},
  \bibnamefont{and} \bibinfo{author}{\bibfnamefont{J.}~\bibnamefont{Tennyson}},
  \bibinfo{journal}{Plasma Sources Science and Technology}
  \textbf{\bibinfo{volume}{29}}, \bibinfo{pages}{10LT01}
  (\bibinfo{year}{2020}).

\bibitem[{\citenamefont{Laporta et~al.}(2012)\citenamefont{Laporta, Celiberto,
  and Wadehra}}]{0963-0252-21-5-055018}
\bibinfo{author}{\bibfnamefont{V.}~\bibnamefont{Laporta}},
  \bibinfo{author}{\bibfnamefont{R.}~\bibnamefont{Celiberto}},
  \bibnamefont{and} \bibinfo{author}{\bibfnamefont{J.~M.}
  \bibnamefont{Wadehra}}, \bibinfo{journal}{Plasma Sources Science and
  Technology} \textbf{\bibinfo{volume}{21}}, \bibinfo{pages}{055018}
  (\bibinfo{year}{2012}).

\bibitem[{\citenamefont{Bardsley and Mandl}(1968)}]{0034-4885-31-2-302}
\bibinfo{author}{\bibfnamefont{J.~N.} \bibnamefont{Bardsley}} \bibnamefont{and}
  \bibinfo{author}{\bibfnamefont{F.}~\bibnamefont{Mandl}},
  \bibinfo{journal}{Reports on Progress in Physics}
  \textbf{\bibinfo{volume}{31}}, \bibinfo{pages}{471} (\bibinfo{year}{1968}).

\bibitem[{\citenamefont{Domcke}(1991)}]{Domcke199197}
\bibinfo{author}{\bibfnamefont{W.}~\bibnamefont{Domcke}},
  \bibinfo{journal}{Physics Reports} \textbf{\bibinfo{volume}{208}},
  \bibinfo{pages}{97 } (\bibinfo{year}{1991}).

\bibitem[{\citenamefont{Dub\'e and Herzenberg}(1979)}]{PhysRevA.20.194}
\bibinfo{author}{\bibfnamefont{L.}~\bibnamefont{Dub\'e}} \bibnamefont{and}
  \bibinfo{author}{\bibfnamefont{A.}~\bibnamefont{Herzenberg}},
  \bibinfo{journal}{Phys. Rev. A} \textbf{\bibinfo{volume}{20}},
  \bibinfo{pages}{194} (\bibinfo{year}{1979}).

\bibitem[{\citenamefont{Laporta et~al.}(2021)\citenamefont{Laporta, Agnello,
  Fubiani, Furno, Hill, Reiter, and Taccogna}}]{Laporta_2021}
\bibinfo{author}{\bibfnamefont{V.}~\bibnamefont{Laporta}},
  \bibinfo{author}{\bibfnamefont{R.}~\bibnamefont{Agnello}},
  \bibinfo{author}{\bibfnamefont{G.}~\bibnamefont{Fubiani}},
  \bibinfo{author}{\bibfnamefont{I.}~\bibnamefont{Furno}},
  \bibinfo{author}{\bibfnamefont{C.}~\bibnamefont{Hill}},
  \bibinfo{author}{\bibfnamefont{D.}~\bibnamefont{Reiter}}, \bibnamefont{and}
  \bibinfo{author}{\bibfnamefont{F.}~\bibnamefont{Taccogna}},
  \bibinfo{journal}{Plasma Physics and Controlled Fusion}
  \textbf{\bibinfo{volume}{63}}, \bibinfo{pages}{085006}
  (\bibinfo{year}{2021}).



\bibitem[{\citenamefont{Celiberto et~al.}(2016)\citenamefont{Celiberto, Baluja,
  Janev, and Laporta}}]{0741-3335-58-1-014024}
\bibinfo{author}{\bibfnamefont{R.}~\bibnamefont{Celiberto}},
  \bibinfo{author}{\bibfnamefont{K.~L.} \bibnamefont{Baluja}},
  \bibinfo{author}{\bibfnamefont{R.~K.} \bibnamefont{Janev}}, \bibnamefont{and}
  \bibinfo{author}{\bibfnamefont{V.}~\bibnamefont{Laporta}},
  \bibinfo{journal}{Plasma Physics and Controlled Fusion}
  \textbf{\bibinfo{volume}{58}}, \bibinfo{pages}{014024}
  (\bibinfo{year}{2016}).






\bibitem[{\citenamefont{Laporta
  et~al.}(2016{\natexlab{a}})\citenamefont{Laporta, Tennyson, and
  Celiberto}}]{0963-0252-25-6-06LT02}
\bibinfo{author}{\bibfnamefont{V.}~\bibnamefont{Laporta}},
  \bibinfo{author}{\bibfnamefont{J.}~\bibnamefont{Tennyson}}, \bibnamefont{and}
  \bibinfo{author}{\bibfnamefont{R.}~\bibnamefont{Celiberto}},
  \bibinfo{journal}{Plasma Sources Science and Technology}
  \textbf{\bibinfo{volume}{25}}, \bibinfo{pages}{06LT02}
  (\bibinfo{year}{2016}{\natexlab{a}}).


\bibitem[{\citenamefont{Laporta et~al.}(2013)\citenamefont{Laporta, Celiberto,
  and Tennyson}}]{0963-0252-22-2-025001}
\bibinfo{author}{\bibfnamefont{V.}~\bibnamefont{Laporta}},
  \bibinfo{author}{\bibfnamefont{R.}~\bibnamefont{Celiberto}},
  \bibnamefont{and} \bibinfo{author}{\bibfnamefont{J.}~\bibnamefont{Tennyson}},
  \bibinfo{journal}{Plasma Sources Science and Technology}
  \textbf{\bibinfo{volume}{22}}, \bibinfo{pages}{025001}
  (\bibinfo{year}{2013}).


\bibitem[{\citenamefont{Allan}(2005)}]{0953-4075_38_5_011}
\bibinfo{author}{\bibfnamefont{M.}~\bibnamefont{Allan}}, \bibinfo{journal}{J.
  Phys. B: At. Mol. Opt. Phys.} \textbf{\bibinfo{volume}{38}},
  \bibinfo{pages}{603} (\bibinfo{year}{2005}).

\bibitem[{\citenamefont{Zhang et~al.}(2004)\citenamefont{Zhang, Vanroose,
  McCurdy, Orel, and Rescigno}}]{PhysRevA.69.062711}
\bibinfo{author}{\bibfnamefont{Z.}~\bibnamefont{Zhang}},
  \bibinfo{author}{\bibfnamefont{W.}~\bibnamefont{Vanroose}},
  \bibinfo{author}{\bibfnamefont{C.~W.} \bibnamefont{McCurdy}},
  \bibinfo{author}{\bibfnamefont{A.~E.} \bibnamefont{Orel}}, \bibnamefont{and}
  \bibinfo{author}{\bibfnamefont{T.~N.} \bibnamefont{Rescigno}},
  \bibinfo{journal}{Phys. Rev. A} \textbf{\bibinfo{volume}{69}},
  \bibinfo{pages}{062711} (\bibinfo{year}{2004}).

\bibitem[{\citenamefont{Trevisan et~al.}(2005)\citenamefont{Trevisan, Houfek,
  Zhang, Orel, McCurdy, and Rescigno}}]{PhysRevA.71.052714}
\bibinfo{author}{\bibfnamefont{C.~S.} \bibnamefont{Trevisan}},
  \bibinfo{author}{\bibfnamefont{K.}~\bibnamefont{Houfek}},
  \bibinfo{author}{\bibfnamefont{Z.}~\bibnamefont{Zhang}},
  \bibinfo{author}{\bibfnamefont{A.~E.} \bibnamefont{Orel}},
  \bibinfo{author}{\bibfnamefont{C.~W.} \bibnamefont{McCurdy}},
  \bibnamefont{and} \bibinfo{author}{\bibfnamefont{T.~N.}
  \bibnamefont{Rescigno}}, \bibinfo{journal}{Phys. Rev. A}
  \textbf{\bibinfo{volume}{71}}, \bibinfo{pages}{052714}
  (\bibinfo{year}{2005}).

\bibitem[{\citenamefont{Mojarrabi et~al.}(1995)\citenamefont{Mojarrabi, Gulley,
  Middleton, Cartwright, Teubner, Buckman, and Brunger}}]{0953-4075-28-3-019}
\bibinfo{author}{\bibfnamefont{B.}~\bibnamefont{Mojarrabi}},
  \bibinfo{author}{\bibfnamefont{R.~J.} \bibnamefont{Gulley}},
  \bibinfo{author}{\bibfnamefont{A.~G.} \bibnamefont{Middleton}},
  \bibinfo{author}{\bibfnamefont{D.~C.} \bibnamefont{Cartwright}},
  \bibinfo{author}{\bibfnamefont{P.~J.~O.} \bibnamefont{Teubner}},
  \bibinfo{author}{\bibfnamefont{S.~J.} \bibnamefont{Buckman}},
  \bibnamefont{and} \bibinfo{author}{\bibfnamefont{M.~J.}
  \bibnamefont{Brunger}}, \bibinfo{journal}{Journal of Physics B: Atomic,
  Molecular and Optical Physics} \textbf{\bibinfo{volume}{28}},
  \bibinfo{pages}{487} (\bibinfo{year}{1995}).

\bibitem[{\citenamefont{Sanche and Michaud}(1984)}]{doi:10.1063/1.447371}
\bibinfo{author}{\bibfnamefont{L.}~\bibnamefont{Sanche}} \bibnamefont{and}
  \bibinfo{author}{\bibfnamefont{M.}~\bibnamefont{Michaud}},
  \bibinfo{journal}{The Journal of Chemical Physics}
  \textbf{\bibinfo{volume}{81}}, \bibinfo{pages}{257} (\bibinfo{year}{1984}).

\bibitem[{\citenamefont{Szmytkowski and Maciag}(1991)}]{Szmytkowski_1991}
\bibinfo{author}{\bibfnamefont{C.}~\bibnamefont{Szmytkowski}} \bibnamefont{and}
  \bibinfo{author}{\bibfnamefont{K.}~\bibnamefont{Maciag}},
  \bibinfo{journal}{Journal of Physics B: Atomic, Molecular and Optical
  Physics} \textbf{\bibinfo{volume}{24}}, \bibinfo{pages}{4273}
  (\bibinfo{year}{1991}).

\bibitem[{\citenamefont{Schwartz et~al.}(1952)\citenamefont{Schwartz, Slawsky,
  and Herzfeld}}]{doi:10.1063/1.1700221}
\bibinfo{author}{\bibfnamefont{R.~N.} \bibnamefont{Schwartz}},
  \bibinfo{author}{\bibfnamefont{Z.~I.} \bibnamefont{Slawsky}},
  \bibnamefont{and} \bibinfo{author}{\bibfnamefont{K.~F.}
  \bibnamefont{Herzfeld}}, \bibinfo{journal}{The Journal of Chemical Physics}
  \textbf{\bibinfo{volume}{20}}, \bibinfo{pages}{1591} (\bibinfo{year}{1952}).

\bibitem[{\citenamefont{Nikitin and Troe}(2008)}]{B715095D}
\bibinfo{author}{\bibfnamefont{E.~E.} \bibnamefont{Nikitin}} \bibnamefont{and}
  \bibinfo{author}{\bibfnamefont{J.}~\bibnamefont{Troe}},
  \bibinfo{journal}{Phys. Chem. Chem. Phys.} \textbf{\bibinfo{volume}{10}},
  \bibinfo{pages}{1483} (\bibinfo{year}{2008}).

\bibitem[{\citenamefont{Laporta and Bruno}(2013)}]{laporta:104319}
\bibinfo{author}{\bibfnamefont{V.}~\bibnamefont{Laporta}} \bibnamefont{and}
  \bibinfo{author}{\bibfnamefont{D.}~\bibnamefont{Bruno}},
  \bibinfo{journal}{The Journal of Chemical Physics}
  \textbf{\bibinfo{volume}{138}}, \bibinfo{eid}{104319}
  (pages~\bibinfo{numpages}{9}) (\bibinfo{year}{2013}).

\bibitem[{\citenamefont{Heritier et~al.}(2014)\citenamefont{Heritier, Jaffe,
  Laporta, and Panesi}}]{doi:10.1063/1.4900508}
\bibinfo{author}{\bibfnamefont{K.~L.} \bibnamefont{Heritier}},
  \bibinfo{author}{\bibfnamefont{R.~L.} \bibnamefont{Jaffe}},
  \bibinfo{author}{\bibfnamefont{V.}~\bibnamefont{Laporta}}, \bibnamefont{and}
  \bibinfo{author}{\bibfnamefont{M.}~\bibnamefont{Panesi}},
  \bibinfo{journal}{The Journal of Chemical Physics}
  \textbf{\bibinfo{volume}{141}}, \bibinfo{pages}{184302}
  (\bibinfo{year}{2014}).

\bibitem[{\citenamefont{Alves et~al.}(2016)\citenamefont{Alves, Coche, Ridenti,
  and Guerra}}]{Alves2016}
\bibinfo{author}{\bibfnamefont{L.~L.} \bibnamefont{Alves}},
  \bibinfo{author}{\bibfnamefont{P.}~\bibnamefont{Coche}},
  \bibinfo{author}{\bibfnamefont{A.~M.} \bibnamefont{Ridenti}},
  \bibnamefont{and} \bibinfo{author}{\bibfnamefont{V.}~\bibnamefont{Guerra}},
  \bibinfo{journal}{The European Physical Journal D}
  \textbf{\bibinfo{volume}{70}}, \bibinfo{pages}{1} (\bibinfo{year}{2016}).

\bibitem[{\citenamefont{Laporta
  et~al.}(2016{\natexlab{b}})\citenamefont{Laporta, Heritier, and
  Panesi}}]{Laporta201644}
\bibinfo{author}{\bibfnamefont{V.}~\bibnamefont{Laporta}},
  \bibinfo{author}{\bibfnamefont{K.}~\bibnamefont{Heritier}}, \bibnamefont{and}
  \bibinfo{author}{\bibfnamefont{M.}~\bibnamefont{Panesi}},
  \bibinfo{journal}{Chemical Physics} \textbf{\bibinfo{volume}{472}},
  \bibinfo{pages}{44 } (\bibinfo{year}{2016}{\natexlab{b}}).




\bibitem[{\citenamefont{Gimelshein et~al.}(0)\citenamefont{Gimelshein, Wysong,
  Fangman, Andrienko, Kunova, Kustova, Morgado, Garbacz, Fossati, and
  Hanquist}}]{doi:10.2514/1.T6462}
\bibinfo{author}{\bibfnamefont{S.~F.} \bibnamefont{Gimelshein}},
  \bibinfo{author}{\bibfnamefont{I.~J.} \bibnamefont{Wysong}},
  \bibinfo{author}{\bibfnamefont{A.~J.} \bibnamefont{Fangman}},
  \bibinfo{author}{\bibfnamefont{D.~A.} \bibnamefont{Andrienko}},
  \bibinfo{author}{\bibfnamefont{O.~V.} \bibnamefont{Kunova}},
  \bibinfo{author}{\bibfnamefont{E.~V.} \bibnamefont{Kustova}},
  \bibinfo{author}{\bibfnamefont{F.}~\bibnamefont{Morgado}},
  \bibinfo{author}{\bibfnamefont{C.}~\bibnamefont{Garbacz}},
  \bibinfo{author}{\bibfnamefont{M.}~\bibnamefont{Fossati}}, \bibnamefont{and}
  \bibinfo{author}{\bibfnamefont{K.~M.} \bibnamefont{Hanquist}},
  \bibinfo{journal}{Journal of Thermophysics and Heat Transfer}
  \textbf{\bibinfo{volume}{0}}, \bibinfo{pages}{1} (\bibinfo{year}{0}),
  \eprint{https://doi.org/10.2514/1.T6462},
  \urlprefix\url{https://doi.org/10.2514/1.T6462}.


\bibitem[{\citenamefont{Robben}(1959)}]{doi:10.1063/1.1730368}
\bibinfo{author}{\bibfnamefont{F.}~\bibnamefont{Robben}}, \bibinfo{journal}{The
  Journal of Chemical Physics} \textbf{\bibinfo{volume}{31}},
  \bibinfo{pages}{420} (\bibinfo{year}{1959}).


\bibitem[{\citenamefont{Landau and Teller}(1963)}]{landauteller}
\bibinfo{author}{\bibfnamefont{L.~D.} \bibnamefont{Landau}} \bibnamefont{and}
  \bibinfo{author}{\bibfnamefont{E.}~\bibnamefont{Teller}},
  \bibinfo{journal}{Phys. Z. Sowjetunion} \textbf{\bibinfo{volume}{10}}
  (\bibinfo{year}{1963}).


\bibitem[{\citenamefont{{Narasinga Rao} and
  Taylor}(1968)}]{NARASINGARAO1968296}
\bibinfo{author}{\bibfnamefont{K.}~\bibnamefont{{Narasinga Rao}}}
  \bibnamefont{and} \bibinfo{author}{\bibfnamefont{R.}~\bibnamefont{Taylor}},
  \bibinfo{journal}{Physics Letters A} \textbf{\bibinfo{volume}{27}},
  \bibinfo{pages}{296} (\bibinfo{year}{1968}).



  
  
\bibitem[{\citenamefont{Hagelaar and Pitchford}(2005)}]{hagelaar2005solving}
\bibinfo{author}{\bibfnamefont{G.}~\bibnamefont{Hagelaar}} \bibnamefont{and}
  \bibinfo{author}{\bibfnamefont{L.}~\bibnamefont{Pitchford}},
  \bibinfo{journal}{Plasma Sources Science and Technology}
  \textbf{\bibinfo{volume}{14}}, \bibinfo{pages}{722} (\bibinfo{year}{2005}).

\bibitem[{Hay()}]{Hayashi}
\emph{\bibinfo{title}{Hayashi database}}, \bibinfo{note}{{www.lxcat.net},
  retrieved on October 13, 2021}.

\bibitem[{\citenamefont{Fowler}(1924)}]{fowler1924xxiii}
\bibinfo{author}{\bibfnamefont{R.~H.} \bibnamefont{Fowler}},
  \bibinfo{journal}{The London, Edinburgh, and Dublin Philosophical Magazine
  and Journal of Science} \textbf{\bibinfo{volume}{47}}, \bibinfo{pages}{257}
  (\bibinfo{year}{1924}).

\bibitem[{\citenamefont{Vialetto et~al.}(2021)\citenamefont{Vialetto, Moussa,
  van Dijk, Longo, Diomede, Guerra, and Alves}}]{vialetto2021effect}
\bibinfo{author}{\bibfnamefont{L.}~\bibnamefont{Vialetto}},
  \bibinfo{author}{\bibfnamefont{A.~B.} \bibnamefont{Moussa}},
  \bibinfo{author}{\bibfnamefont{J.}~\bibnamefont{van Dijk}},
  \bibinfo{author}{\bibfnamefont{S.}~\bibnamefont{Longo}},
  \bibinfo{author}{\bibfnamefont{P.}~\bibnamefont{Diomede}},
  \bibinfo{author}{\bibfnamefont{V.}~\bibnamefont{Guerra}}, \bibnamefont{and}
  \bibinfo{author}{\bibfnamefont{L.~L.} \bibnamefont{Alves}},
  \bibinfo{journal}{Plasma Sources Sci. Technol.} \textbf{\bibinfo{volume}{30}}
  (\bibinfo{year}{2021}).

\bibitem[{\citenamefont{Tejero{-}del{-}Caz
  et~al.}(2019)\citenamefont{Tejero{-}del{-}Caz, Guerra, Gon{\c{c}}alves,
  da~Silva, Marques, Pinh{$\tilde{a}$}o, Pintassilgo, and
  Alves}}]{tejero2019lisbon}
\bibinfo{author}{\bibfnamefont{A.}~\bibnamefont{Tejero{-}del{-}Caz}},
  \bibinfo{author}{\bibfnamefont{V.}~\bibnamefont{Guerra}},
  \bibinfo{author}{\bibfnamefont{D.}~\bibnamefont{Gon{\c{c}}alves}},
  \bibinfo{author}{\bibfnamefont{M.~L.} \bibnamefont{da~Silva}},
  \bibinfo{author}{\bibfnamefont{L.}~\bibnamefont{Marques}},
  \bibinfo{author}{\bibfnamefont{N.}~\bibnamefont{Pinh{$\tilde{a}$}o}},
  \bibinfo{author}{\bibfnamefont{C.~D.} \bibnamefont{Pintassilgo}},
  \bibnamefont{and} \bibinfo{author}{\bibfnamefont{L.~L.} \bibnamefont{Alves}},
  \bibinfo{journal}{Plasma Sources Sci. Technol.} \textbf{\bibinfo{volume}{28}}
  (\bibinfo{year}{2019}).

\bibitem[{\citenamefont{Takeuchi and
  Nakamura}(2001)}]{takeuchi2001measurements}
\bibinfo{author}{\bibfnamefont{T.}~\bibnamefont{Takeuchi}} \bibnamefont{and}
  \bibinfo{author}{\bibfnamefont{Y.}~\bibnamefont{Nakamura}},
  \bibinfo{journal}{IEEJ Trans.} \textbf{\bibinfo{volume}{121}},
  \bibinfo{pages}{481} (\bibinfo{year}{2001}).

\bibitem[{\citenamefont{Mechlinska-Drewko
  et~al.}(1999)\citenamefont{Mechlinska-Drewko, Roznerski, Petrovic, and
  Karwasz}}]{mechlinska1999dt}
\bibinfo{author}{\bibfnamefont{J.}~\bibnamefont{Mechlinska-Drewko}},
  \bibinfo{author}{\bibfnamefont{W.}~\bibnamefont{Roznerski}},
  \bibinfo{author}{\bibfnamefont{Z.~L.} \bibnamefont{Petrovic}},
  \bibnamefont{and} \bibinfo{author}{\bibfnamefont{G.~P.}
  \bibnamefont{Karwasz}}, \bibinfo{journal}{J. Phys. D: Appl. Phys.}
  \textbf{\bibinfo{volume}{32}}, \bibinfo{pages}{2746} (\bibinfo{year}{1999}).

\bibitem[{\citenamefont{Petrovi{\'c} et~al.}(2009)\citenamefont{Petrovi{\'c},
  Dujko, Mari{\'c}, Malovi{\'c}, Nikitovi{\'c}, {\v{S}}a{\v{s}}i{\'c},
  Jovanovi{\'c}, Stojanovi{\'c}, and
  Radmilovi{\'c}-Radjenovi{\'c}}}]{petrovic2009measurement}
\bibinfo{author}{\bibfnamefont{Z.~L.} \bibnamefont{Petrovi{\'c}}},
  \bibinfo{author}{\bibfnamefont{S.}~\bibnamefont{Dujko}},
  \bibinfo{author}{\bibfnamefont{D.}~\bibnamefont{Mari{\'c}}},
  \bibinfo{author}{\bibfnamefont{G.}~\bibnamefont{Malovi{\'c}}},
  \bibinfo{author}{\bibfnamefont{{\v{Z.}}.}~\bibnamefont{Nikitovi{\'c}}},
  \bibinfo{author}{\bibfnamefont{O.}~\bibnamefont{{\v{S}}a{\v{s}}i{\'c}}},
  \bibinfo{author}{\bibfnamefont{J.}~\bibnamefont{Jovanovi{\'c}}},
  \bibinfo{author}{\bibfnamefont{V.}~\bibnamefont{Stojanovi{\'c}}},
  \bibnamefont{and}
  \bibinfo{author}{\bibfnamefont{M.}~\bibnamefont{Radmilovi{\'c}-Radjenovi{\'c}}},
  \bibinfo{journal}{J. Phys. D: Appl. Phys.} \textbf{\bibinfo{volume}{42}}
  (\bibinfo{year}{2009}).

\bibitem[{\citenamefont{Stokes et~al.}(2021)\citenamefont{Stokes, White,
  Campbell, and Brunger}}]{stokes2021toward}
\bibinfo{author}{\bibfnamefont{P.~W.} \bibnamefont{Stokes}},
  \bibinfo{author}{\bibfnamefont{R.~D.} \bibnamefont{White}},
  \bibinfo{author}{\bibfnamefont{L.}~\bibnamefont{Campbell}}, \bibnamefont{and}
  \bibinfo{author}{\bibfnamefont{M.~J.} \bibnamefont{Brunger}},
  \bibinfo{journal}{J. Chem. Phys.} \textbf{\bibinfo{volume}{155}}
  (\bibinfo{year}{2021}).

\bibitem[{\citenamefont{Josic et~al.}(2001)\citenamefont{Josic, Wr{\'o}blewski,
  Petrovic, Mechlinska-Drewko, and Karwasz}}]{Josic2001318}
\bibinfo{author}{\bibfnamefont{L.}~\bibnamefont{Josic}},
  \bibinfo{author}{\bibfnamefont{T.}~\bibnamefont{Wr{\'o}blewski}},
  \bibinfo{author}{\bibfnamefont{Z.~L.} \bibnamefont{Petrovic}},
  \bibinfo{author}{\bibfnamefont{J.}~\bibnamefont{Mechlinska-Drewko}},
  \bibnamefont{and} \bibinfo{author}{\bibfnamefont{G.~P.}
  \bibnamefont{Karwasz}}, \bibinfo{journal}{Chemical Physics Letters}
  \textbf{\bibinfo{volume}{350}}, \bibinfo{pages}{318 } (\bibinfo{year}{2001}).



\bibitem[{LxC()}]{LxCat_Laporta}
\emph{\bibinfo{title}{Laporta database}},
  \bibinfo{note}{{www.lxcat.net/Laporta}, retrieved on October 13, 2021}.

\end{thebibliography}

\end{document}